\documentclass[11pt,a4paper]{article}
\usepackage{jheppub}
\usepackage{bm}
\usepackage{amsmath}	

\usepackage{comment}

\usepackage{lineno,hyperref}
\usepackage{amsmath}
\usepackage{amssymb}

\usepackage{listings}

\definecolor{codegreen}{rgb}{0,0.6,0}
\definecolor{codegray}{rgb}{0.5,0.5,0.5}
\definecolor{codepurple}{rgb}{0.58,0,0.82}
\definecolor{backcolour}{rgb}{0.95,0.95,0.92}

\lstdefinestyle{mystyle}{
    backgroundcolor=\color{backcolour},   
    commentstyle=\color{codegreen},
    keywordstyle=\color{magenta},
    numberstyle=\tiny\color{codegray},
    stringstyle=\color{codepurple},
    basicstyle=\ttfamily\footnotesize,
    breakatwhitespace=false,         
    breaklines=true,                 
    captionpos=b,                    
    keepspaces=true,                 
    numbersep=5pt,                  
    showspaces=false,                
    showstringspaces=false,
    showtabs=false,                  
    tabsize=2
}

\modulolinenumbers[5]

\newcommand\MSbar{\ensuremath{\overline{\mathrm{MS}}}}
\newcommand\FORM{\texttt{FORM} }
\newcommand\Mathematica{\texttt{Mathematica}}
\newcommand\fermat{\texttt{Fermat}}
\newcommand\hc{\ensuremath{\mathrm{h.c.}}}
\newcommand\Macaulay{\texttt{Macaulay2}}
\newcommand\LT{\ensuremath{\mathrm{LT}}}

\newcommand\tr[1]{\ensuremath{\mathrm{tr}#1}}
\newcommand\myRe{\ensuremath{\mathrm{Re}}}
\newcommand\myIm{\ensuremath{\mathrm{Im}}}
\newcommand\lamS{\ensuremath{\Lambda_{00}}}
\newcommand\mS{\ensuremath{M_0}}
\newcommand\lamV{\ensuremath{\vec{\Lambda}}}
\newcommand\mV{\ensuremath{\vec{M}}}
\newcommand\lamT{\ensuremath{{\Lambda}}}
\newcommand\trlam[2]{\ensuremath{\mathrm{tr}^{#1}\Lambda^{#2}}}

\newcommand\T[3]{\ifodd \the\numexpr #2+#3 
		\textcolor{red}{\ensuremath{T_{#1#2#3}}} 
   \else 
	 \ifnum#2=#3
		 \ifnum#2=1
			 \ifnum#1>0
				\textcolor{blue}{\ensuremath{T_{#1#2#3}}} 
			\else
				\ensuremath{T_{#1#2#3}} 
			\fi
		 \else
			\ensuremath{T_{#1#2#3}} 
		 \fi
	 \else
		 \ensuremath{T_{#1#2#3}} 
	\fi
   \fi}

\newcommand\Z[2]{\ensuremath{Z_{#1_{(#2)}}}}
\newcommand\Y[1]{\ensuremath{Y_1}}

\newcommand\IdealTHDM{\ensuremath{\mathcal{I}_{\text{2HDM}}}}
\newcommand\ModuleTHDM{\ensuremath{\mathcal{M}_{\text{2HDM}}}}
\title{On the scalar sector of 2HDM: ring of basis invariants, syzygies, and six-loop renormalization-group equations}

\author{A.V. Bednyakov}
\emailAdd{alexander.bednyakov@jinr.ru}

\affiliation{Bogoliubov Laboratory of Theoretical Physics, Joint Institute for Nuclear Research, \\ 
Joliot-Curie, 6, Dubna 141980, Russia}

\abstract{
	We consider a generating set of reparametrization invariants that can be constructed from the couplings and masses entering the scalar potential of the general Two-Higgs-Doublet Model (2HDM). 
	Being independent of higgs-basis rotations, they generate a polynomial ring of basis invariants that represent the physical content of the model. 
	Ignoring for the moment gauge and Yukawa interactions, we derive six-loop renormalization group equations (RGE) for all the invariants entering the set.  
	We do not compute a single Feynman diagram but rely heavily on the general RGE results for scalar theories. 
	We use linear algebra together with techniques from Invariant Theory.
	The latter not only allow one to compute the number of linearly independent invariants entering beta functions at a certain loop order (via Hilbert series) 
	but also provide a convenient tool for dealing with polynomial relations (so-called syzygies) between invariants from the generating set.  
}


\begin{document}
\maketitle

\section{\label{sec:intro}Introduction}

	Quantum field theory (QFT) provides very accurate predictions for a plethora of observables 
	expressing the latter in terms of a number of parameters, 
	which, when considered in the context of perturbation theory (PT), include also an auxiliary renormalization scale $\mu$. 

	A convenient tool for  dealing with high-order terms in PT expansion is dimensional regularization \cite{tHooft:1972tcz} 
	accompanied by modified minimal ($\MSbar$) subtractions. 
	In the \MSbar\ scheme  QFT model parameters depend implicitly on the scale $\mu$ and the dependence is governed by differential renormalization group equations (RGE). 
	While the boundary conditions should be extracted from experiment, the RG functions (beta functions and anomalous dimensions) can be calculated order-by-order in PT.
	The  RGE solution allows improving the precision of finite-order predictions by resumming certain logarithmic corrections into the redefinition of model parameters. 

	However, physical questions, e.g., whether the CP symmetry is broken or not, can be obscured if there are basis redundancies. 
	A well-known example is the CP non-conservation in the weak sector of the Standard Model (SM). 
	In spite of the fact that the general SM Lagrangian involves many complex phases that enter quark Yukawa matrices, a convenient  choice  of basis
	can reduce their number to a single physical phase. It is obvious that predictions should not depend on reparametrization and  
	an alternative, basis-independent, formulation of the CP conservation condition is possible, i.e., the vanishing of the Jarlskog invariant \cite{Jarlskog:1985ht}. 

	Let us also mention that basis redundancies being related to global symmetries of the kinetic terms in the QFT Lagrangians can lead to ambiguities in the RG equations for the Lagrangian parameters (see, e.g., refs.~\cite{Bednyakov:2012en,Herren:2017uxn} for explicit three-loop computations, and refs.~\cite{Jack:2013sha,Herren:2021yur,Davies:2021mnc} for a general discussion). On the contrary, one can prove that RGEs for basis invariant combinations of the parameters are free of these ambiguities. 

	In this paper we consider the scalar sector of the general (type III) Two-Higgs-Doublet Model (2HDM) (for review see refs.~\cite{Branco:2011iw,Ivanov:2017dad}). 
	The 2HDM model, being one of the simplest (yet renormalizable) alternatives of the SM, 
	predicts new scalar states in the spectrum --- two neutral $H,A$ and one charged $H^\pm$ higgs bosons. 
	Being linear combinations of components of the SU(2) doublets, their interactions with vector fields are fixed by the postulated gauge symmetry, 
	but there is freedom in self-interactions and fermion Yukawa couplings opening up the possibility for additional sources of CP violation.

	Our current study addresses the issues related, first, to the independence of the observables on reparametrization~(see, e.g., refs.~
	\cite{Botella:1994cs,Davidson:2005cw,Jenkins:2009dy,Hanany:2010vu,Trautner:2018ipq,Bento:2020jei,Bento:2023owf}) and, 
	second, to the renormalization-scale dependence of reparametrization invariants at high orders of PT and thus 
	extends the results of refs.~\cite{Ivanov:2005hg,Bednyakov:2018cmx}.

	In recent years, remarkable progress has been achieved in computation of high-order corrections to the renormalization-group equations, especially in $\phi^4$ \cite{Kompaniets:2017yct,Schnetz:2022nsc} and QCD-like theories (see, ref.~\cite{Baikov:2016tgj,Chetyrkin:2017bjc,Luthe:2017ttg} and references therein). 
Many perturbative calculations in particular models (see, e.g.,  refs.~\cite{Mihaila:2012fm,Mihaila:2012pz,Bednyakov:2012en,Chetyrkin:2013wya,Bednyakov:2014pia,Bednyakov:2015ooa,Zoller:2015tha,Chetyrkin:2016ruf,Herren:2017uxn,Bednyakov:2018cmx}) combined with the consequences of the so-called $a$-theorem  (see, e.g, refs~\cite{Jack:2013sha,Poole:2019txl}) 
	were used to fix the coefficients in the general anzatz for beta functions in an arbitrary renormalizable four-dimensional QFT \cite{Poole:2019kcm,Steudtner:2024teg}. 
	Thanks to modern computer packages \cite{Litim:2020jvl,Thomsen:2021ncy}, current state-of-the-art results \cite{Bednyakov:2021qxa,Davies:2021mnc,Steudtner:2024teg} 
	allow one to obtain gauge-coupling beta functions at four loops and Yukawa and scalar self-coupling beta functions at three loops without calculating a single Feynman diagram.

	All such software deals with Lagrangian parameters. 
	If one is interested in the corresponding basis-invariant combinations and the scale dependence of the latter, one has to make an additional effort\footnote{At the moment, we do not see a way to derive such kind of RGEs directly in terms of invariants.}.
	One of the main results of the current study is the derivation of six-loop RGE for an arbitrary basis invariant constructed from the parameters entering the scalar potential of 2HDM.
	As in ref.~\cite{Bednyakov:2018cmx}, we restrict ourselves to the limit of vanishing gauge and Yukawa interactions.  
	Our computation relies on general six-loop expressions \cite{Bednyakov:2021ojn} obtained in scalar theories thanks to tedious calculations of ref.~\cite{Kompaniets:2017yct}.  
	The approach also incorporates ideas from invariant theory, ring theory and theory of modules (see, e.g, textbooks~\cite{CoxLittleOShea,DerksenKemper} and refs.~\cite{Trautner:2018ipq,Bento:2021hyo}).

The paper is organized as follows. In Section \ref{sec:potential} we review the 2HDM Higgs potential and discuss various parametrizations of the Higgs sector.
In Section~\ref{sec:polynomial_ring} we introduce a polynomial ring of invariants constructed from the parameters of the potential. The variables of the ring correspond to the so-called generating set introduced for the first time in ref.~\cite{Trautner:2018ipq}.  In Sec.~\ref{sec:generating_set} we give a convenient parametrization of the latter in terms of SO(3) covariants. We present a minimal set of relations (syzygies) among the ring variables that generate all other syzygies or, equivalently, a syzygy module in Section~\ref{sec:syzigies}. 
Section~\ref{sec:RG} is devoted to the calculation of six-loop renormalization group equations for the members of the invariant ring. Contrary to the previous studies \cite{Bednyakov:2018cmx}, we do not compute Feynman diagrams but use the algorithm presented in Sec.~\ref{sec:RG_algorithm}. One-loop analytical expressions for the RG functions can be found in Sec.~\ref{sec:RG_results_1l}, while the higher-order results are provided as the supplementary material.  Our conclusions can be found in Sec.~\ref{sec:conclusios}. In a series of appendices, we give brief information on polynomial rings, ideals and Gr\"obner bases (\ref{app:monomial_ordering_and_gb}), and details on the computation of syzygies (\ref{app:gb_syzygies}) via the \Macaulay\ package \cite{M2}. Appendix~\ref{app:Hilbert_series} is devoted to a free resolution of the syzygy module and its relation to the Hilbert series that counts basis invariants. Finally, we provide polynomial equations that express an invariant from the generating set via algebraically independent invariants introduced in ref.~\cite{Trautner:2018ipq}.

	\section{\label{sec:potential}The scalar potential of 2HDM}
	The most general renormalizable Higgs potential can be written in the following form:
\begin{align}
	V_H & = 
		  m_{11}^2 \Phi_1^\dagger \Phi_1
		+ m_{22}^2 \Phi_2^\dagger \Phi_2
		- \left(m_{12}^2 \Phi_1^\dagger \Phi_2 + \hc\right)
		\nonumber\\
		& 
		+ \frac{1}{2} \lambda_1 \left(\Phi_1^\dagger \Phi_1\right)^2 
		+ \frac{1}{2} \lambda_2 \left(\Phi_2^\dagger \Phi_2\right)^2 
		+  \lambda_3 \left(\Phi_1^\dagger \Phi_1\right)\left(\Phi_2^\dagger \Phi_2\right) 
		+  \lambda_4 \left(\Phi_1^\dagger \Phi_2\right)\left(\Phi_2^\dagger \Phi_1\right) 
		\nonumber\\
		&+ \left[ 
			\frac{1}{2} \lambda_5 \left(\Phi_1^\dagger \Phi_2\right)^2
			+ \lambda_6 \left(\Phi_1^\dagger \Phi_1\right) \left(\Phi_1^\dagger \Phi_2\right)
			+ \lambda_7 \left(\Phi_2^\dagger \Phi_2\right) \left(\Phi_1^\dagger \Phi_2\right)
			+ \hc
		\right]
\label{eq:V_gen}
\end{align}
where $\Phi_{1,2}$ are the $SU(2)$ doublets. The self-couplings $\lambda_{1-4}$ and the mass parameters  $m_{11}^2$, $m_{22}^2$ are real, while $\lambda_{5-7}$, and $m_{12}^2$ can be complex. Not all of the fourteen real parameters are physical due to the freedom in redefining the Higgs basis by the unitary rotation
\begin{align}
	\Phi_a \to U_{ab} \Phi_b, \quad U \in U(2),
\end{align}
where $a,b=1,2$ enumerate the doublets. One can see that the overall phase of $U$ does not change the couplings and masses in \eqref{eq:V_gen}, so in what follows we restrict ourselves to $U\in SU(2)$. The three parameters of the $SU(2)$ rotation can be used to get rid of three out of 14 parameters of the potential and, thus, we are left only with $N=11$ independent quantities. 

There is an alternative notation \cite{Botella:1994cs}
\begin{align}
	V_H & = \frac{1}{2} \lambda_{ab,cd} (\Phi^\dagger_a \Phi_b)(\Phi^\dagger_c \Phi_d) 
	+ m^2_{ab} (\Phi^\dagger_a \Phi_b),\quad \lambda_{ab,cd} = \lambda_{cd,ba}, \quad
	m^2_{ba} = m^{\dagger2}_{ab},
\label{eq:V_m_lambda}
\end{align}
which can be used as an intermediate step to rewrite the self-couplings (see refs.~\cite{Branco:2011iw,Ivanov:2017dad}) 
\begin{align}
	\lambda_{ab,cd} & = \frac{1}{2} \Lambda_{\mu\nu} \sigma^\mu_{ab} \sigma^\nu_{cd} 
	= \frac{1}{2} \left[
		\lamS \delta_{ab} \delta_{cd} + \lamV \left( \vec{\sigma}_{ab} \delta_{cd}
		+ \delta_{ab} \vec{\sigma}_{cd} \right)
		+  \vec{\sigma}_{ab} \cdot  \Lambda_{ij} \cdot \vec{\sigma}_{cd} 
	\right],
	\label{eq:lambda_to_Lambda}
\end{align}
\begin{align}
	\Lambda_{\mu\nu} & = \frac{1}{2} \lambda_{ab,cd} \sigma_\mu^{ba} \sigma_\nu^{dc}
	= 
	\begin{pmatrix}
		\frac{\lambda_1 + \lambda_2}{2} + \lambda_3 & 
	 \myRe\left(\lambda_6 + \lambda_7\right) & 
	-\myIm\left(\lambda_6 + \lambda_7\right) &
		\frac{\lambda_1 - \lambda_2}{2} \\
		\phantom{-}\myRe\left(\lambda_6 + \lambda_7\right) &
		\lambda_4 + \myRe \left( \lambda_5\right) &
		- \myIm \left( \lambda_5 \right) &
		\phantom{-}\myRe\left(\lambda_6 - \lambda_7\right)
	\\
	- \myIm \left( \lambda_6 + \lambda_7\right)&
	- \myIm \left( \lambda_5 \right) &
	\lambda_4 - \myRe \left( \lambda_5\right) &
	- \myIm \left( \lambda_6 - \lambda_7\right) \\
	\frac{\lambda_1 - \lambda_2}{2} &
	\myRe\left( \lambda_6 - \lambda_7\right) &
	- \myIm \left( \lambda_6 - \lambda_7 \right) &
	\frac{\lambda_1 + \lambda_2}{2} - \lambda_3
	\end{pmatrix}
\label{eq:LambdaMuNudef}
\end{align}
in terms of a scalar $\lamS$, a vector $\lamV$, and a symmetric matrix $\lamT$, where $\sigma^\mu \equiv (1,\vec{\sigma})$, 
	$\mu,\nu=0,1,2,3$, $i,j=1,2,3$ and the euclidean metric is used for both four- and three-dimensional indices. 
The same trick can be used for the mass term:
\begin{align}
	m^2_{ab} & = \frac{1}{2} M_\mu \sigma^\mu_{ab} = \frac{1}{2} \left[ M_0 \delta_{ab} + \vec{M} \vec{\sigma}_{ab} \right],
	\quad
\label{eq:m_to_M} 
\end{align}
\begin{align}
\mS & = \tr\left[m^2\right] = m^2_{11} + m_{22}^2, 
\quad \mV = \tr \left[m^2 \vec{\sigma} \right] = \left( 
		- 2\myRe\, m_{12}^2,
		  2\myIm\, m_{12}^2,
	  m_{11}^2 - m_{22}^2
  \right).
\end{align}
We can also decompose the tensor
\begin{align}
	\Phi_a \Phi^\dagger_b & = \frac{1}{2} \left(\Phi^\dagger \Phi\right) \delta_{ab}
	+ \frac{1}{2} \left(\Phi^\dagger \sigma^n \Phi\right) \sigma^n_{ab}
	= r_0 \, \delta_{ab} + \vec{r} \cdot \vec{\sigma}_{ab} = r_\mu \sigma^\mu_{ab}
	\label{eq:PhiPhi_decomposition}
\end{align}
in terms of a singlet $r_0$ and a vector $\vec{r}$. 
By means of Eqs.\eqref{eq:lambda_to_Lambda},\eqref{eq:m_to_M}, and \eqref{eq:PhiPhi_decomposition} one can rewrite the potential \eqref{eq:V_m_lambda} as 
\begin{align}
	V_H = \frac{1}{4} \Lambda_{\mu\nu} 
	  r_\rho 
	 r_{\sigma} 
	 \left[
	\sigma^\mu_{ab} \sigma^\rho_{ba}
	\sigma^\nu_{cd} \sigma^\sigma_{dc}
\right]
	+ \frac{1}{2} M_\mu r_\nu 
	\left[
	   \sigma^\mu_{ab}
	   \sigma^\nu_{ba}
   	\right]
	=
	M_\mu r^\mu + \Lambda_{\mu\nu} r^\mu r^\nu
	\label{eq:V_M_Lambda}
\end{align}

Under a Higgs-basis change $\Phi_a \to U_{ab} \Phi_b$, $U_{ab} \in SU(2)$, 
$r_0$, $\lamS$, and $\mS$ transform as singlets, while $\vec{r}$, $\lamV$, and $\mV$ transform as triplets under the corresponding $SO(3)$ rotation
\begin{align}
	R_{ij}(U) &  = \frac{1}{2} \tr\left[ U^\dagger \sigma_i U \sigma_j \right], \qquad \det R = + 1.
	\label{eq:proper_rotation}
\end{align}
The symmetric $3\times3$ matrix $\lamT\equiv\{\Lambda_{ij}\}$ can be decomposed into a singlet $\trlam{}{}$ and
a five-plet $\tilde \Lambda_{ij} \equiv \left[\Lambda_{ij} - \frac{1}{3} \trlam{}{} \delta_{ij}\right]$.  
In what follows, we consider basis-independent quantities constructed from the covariants $\lamV$, $\mV$, and $\Lambda_{ij}$.

Let us also mention that a given general CP transformation \cite{Branco:2011iw} 
\begin{align}
	\Phi_a \to X_{ab} \Phi_b^*, \qquad X \in U(2) 
	\label{eq:GCP_2x2}
\end{align}
can be represented as the improper rotation
\begin{align}
	\tilde R_{ij}(X) = \frac{1}{2}\tr\left[ X^\dagger \sigma_i X \sigma^T_j\right], \qquad \det \tilde R = -1.
	\label{eq:improper_rotation}
\end{align}
Due to this, one can split the basis-invariant quantities into two groups: CP-even and CP-odd, depending on whether the number of triplets $\lamV$ or $\mV$ entering an invariant is even or odd\footnote{From a geometrical point of view an odd number of 3d vectors can be arranged into a rotation invariant only by means of a triple product.}  \cite{Trautner:2018ipq}. As a consequence, CP-odd invariants change the sign under improper rotations (general CP transformation).

\section{\label{sec:polynomial_ring}Ring of 2HDM basis invariants as a graded polynomial ring}

The reparametrization invariants form a ring $\mathcal{R}$, since sums or products of invariants also inherit this property. An important question is how many \emph{independent } invariants we can construct from the parameters of the Higgs potential. From physical considerations it should be equal to the number $N$ of parameters that can not be set to zero by a basis rotation. In 2HDM one can encode these physical parameters in a set of $N=11$ \emph{algebraically independent} basis invariants $\{f_i\}$, $i=1\ldots N$.

Any other basis-invariant combination of parameters entering the tree-level potential \eqref{eq:V_gen}, say $\mathcal{J}$, can be represented as an \emph{algebraic function} of the above-mentioned independent invariants $f_i$, i.e.,  there exists a polynomial relation $P$ between $\mathcal{J}$ and $\{f_i\}$ 
\begin{align}
	P(\mathcal{J},f_1, \ldots, f_{N}) = 0, \qquad [N=11~\text{for 2HDM}].
\label{eq:algebric_relation}
\end{align}

However, it is more convenient \cite{Trautner:2018ipq} to extend the set of \emph{algebraically independent} invariants to a \emph{generating set} $\{g_n\}$, with $n=1\ldots M$ and $N<M$. 
The advantage of the generating set is that one can express \emph{any} invariant $\mathcal{J}$ as \emph{a polynomial} $\mathcal P$ in $g_n$:
\begin{align*}
	\mathcal{J} = \mathcal P(g_1, \ldots, g_{M}) \qquad [M=22~\text{for 2HDM}]. 
\end{align*}

Due to this, we can represent $\mathcal{R}$ as a polynomial ring
$R \equiv \mathtt{k}[g_1,\ldots,g_M]$ over reals\footnote{In our calculations we restrict ourselves to polynomial rings over rationals $\mathtt{k}=\mathbb{Q}$ or over a ring $\mathtt{k} = \mathbb{Q}[\zeta_3,\ldots,\zeta_9, \zeta_{3,5}]$ generated by (multiple)zeta values appearing in the six-loop RGE for scalar theories \cite{Kompaniets:2017yct,Bednyakov:2021ojn}.}
$\mathtt{k}=\mathbb{R}$
in $M$ unknowns, which we also denote as $\{g_n\}$. The map $\phi$ substitutes the indeterminates by their expressions in terms of the couplings and masses.

Since not all of $g_n$ are algebraically independent, there should be \emph{polynomial relations} 
 between them. It is easy to convince oneself that 
 the latter form an \emph{ideal} $\IdealTHDM = \ker \phi$ in the polynomial ring $R$
 : 
 given any two such relations $P_{1,2}(g_1\ldots g_{M}) = 0$, one can show that $p_1 P_1 + p_2 P_2 = 0$ for any $p_{1,2} \in R$. 
 This subtlety explains why the invariant ring $\mathcal{R}=\mathtt{k}[g_1,\ldots, g_M]/\IdealTHDM$  is actually a quotient ring with an equivalence relation for $p_{1,2}\in R$: $p_1 \sim p_2$ if and only if $p_1 - p_2 \in \IdealTHDM$.  

We use Gr\"obner-basis  methods (see, e.g., ref.~\cite{CoxLittleOShea} and appendices~\ref{app:monomial_ordering_and_gb} and \ref{app:gb_syzygies}) to find a minimal set of relations $\{r_k\}$, 
or \emph{first syzygies} that \emph{generate} the ideal $\IdealTHDM$, i.e., a set of $K$ polynomials in $M$ variables that simultaneously equal zero:
\begin{align}
	r_k(g_1,\ldots,g_{M}) \stackrel{\phi}{\longrightarrow}  0, \qquad k = 1\ldots K \qquad [K = 63~\text{for 2HDM}]. 
\label{eq:def_first_syzygies}
\end{align}
One writes $\IdealTHDM = \langle r_1, \ldots, r_{63}\rangle$, and any element (relation) from $\IdealTHDM$ can be represented as a \emph{linear} combination of the generators $r_k$ with the coefficients from  the ring $R$. 
Given a Gr\"obner basis for $\IdealTHDM$, one can reduce elements from $\mathcal{R}$ to its \emph{normal form} (see appendix \ref{app:monomial_ordering_and_gb}) after arithmetic operations.

One important property of the invariant ring is its grading, i.e., it can be decomposed into a direct sum as
\begin{align}
	\mathcal{R} = \bigoplus_{n=0}^\infty \mathcal{R}_n = \mathcal{R}_0 \oplus \mathcal{R}_1 \oplus \mathcal{R}_2 \oplus \cdots \qquad \mathcal{R}_i \mathcal{R}_j \subseteq \mathcal{R}_{i+j}   
	\label{eq:ring_grading_def}
\end{align}
We call a non-zero element of $\mathcal{R}_n$ a homogeneous element of degree $n$. 
When dealing with the scalar sector of 2HDM,  we associate the index $n$ in \eqref{eq:ring_grading_def} with the total number of SO(3) representations used to construct an invariant. For example, $\lamS, \mS, \trlam{}{} \in \mathcal{R}_1$ are degree-one invariants, while $\lamV\cdot\lamV $ is an example of degree-two homogeneous element belonging to $\mathcal{R}_2$. Moreover, $\lamS (\lamV\cdot \lamV) \in \mathcal{R}_3$, etc.  Perturbative expansion of a basis invariant quantity (e.g., a beta function of basis invariant) is an example of non-homogeneous element belonging to $\mathcal{R}$. However, it can be represented as a sum over homogeneous components. 

A convenient tool to count \emph{linear independent elements}\footnote{Here, linear independence is assumed over $\mathtt{k}$.} of $\mathcal{R}_n$, i.e. $\dim \mathcal{R}_n$,  is the Hilbert series (see, e.g., refs.~\cite{Benvenuti:2006qr,Feng:2007ur,Hanany:2008sb,Lehman:2015via,Gripaios:2020hya,Wang:2021wdq,Yu:2021cco,Yu:2022ttm} for the introduction and various applications)
\begin{align*}
	H(t) = \sum\limits_{n=0}^\infty (\dim  \mathcal{R}_n) t^n, \qquad \dim \mathcal{R}_0 = 1
\end{align*}
Before proceeding further, let us mention that one can treat the index $n$ in \eqref{eq:ring_grading_def} as a multiindex. 
In ref.~\cite{Trautner:2018ipq}, a multigrading is used for a subring $\tilde{\mathcal{ R}}$ of $\mathcal{R}$ from which one excludes trivial degree-one invariants corresponding to $\lamS$, $\mS$ and $\trlam{}{}$. 
Following ref.~\cite{Trautner:2018ipq}, we 
also use the notation 
$[abc]$ (or $[a,b,c]$) to represent the multidegree of an invariant $T_{abc}$ 
built from $a$ copies of the five-plet $\tilde \lamT$, $b$ copies of the triplet $\mV$, and $c$ copies of the triplet $\lamV$. Due to this grading, one has the following scaling:    
\begin{align}
	T_{abc} \to x^a y^b z^c T_{abc}, \qquad \text{if } \tilde\lamT \to x \cdot \tilde\lamT, \quad \mV \to y \cdot \mV, \quad \lamV \to z \cdot \lamV 
\label{eq:a_b_c_scaling}
\end{align}
allowing one to confirm that multidegree of the product $T_{a_1 b_1 c_1} T_{a_2 b_2 c_3}$
is $[a_1+a_2,b_1+b_2,c_1+c_2]$. Given such a grading, one can construct a multigraded Hilbert series that depends on three variables, named in ref.~\cite{Trautner:2018ipq} $y$, $t$, and $q$:
\begin{align}
	H(q,y,t) \equiv \sum\limits_{a,b,c} (\dim \mathcal{\tilde R}_{abc}) q^a y^b t^c.
	\label{eq:HS_q_y_t}
\end{align}
Here $\mathcal{\tilde{R}}_{abc}$ consists of all invariants from $\mathcal{\tilde R}$ having multidegree $[abc]$, and the coefficient of $q^a y^b t^c$ in Taylor expansion of $\mathcal{H}(q,y,t)$ counts linear independent elements of $\mathcal{\tilde R}_{abc}$. 

In this paper, we aim to compute  RG functions in the $\MSbar$ scheme and introduce 
 another multigrading allowing one to count invariants that can appear at different orders of perturbation theory.  
 We use $\{\alpha,\beta\}$ to denote the multidegree of an invariant $I_{\alpha,\beta}$ that scales as\footnote{Note that in ref.~\cite{Bednyakov:2018cmx} we also use 
the notation  $I_{n,m}$ for a particular set of invariants, but with different meaning. While in ref.~\cite{Bednyakov:2018cmx} $n$ still counts the powers of self couplings, the index $m$ has nothing to do with the mass parameter.}
\begin{align}
	I_{\alpha,\beta} \to x^\alpha y^\beta I_{\alpha,\beta},\qquad \text{if } \lambda_i \to x \cdot \lambda_i, \quad m^2_{ij} \to y \cdot m^2_{ij}
\label{eq:n_m_grading}
\end{align}
Obviously, with an invariant $T_{abc}$ having multidegree $[a,b,c]$ we can associate the multidegree $\{a+c,b\}$. Moreover, the singlets 
$\lamS$, $\trlam{}{}$, and $\mS$ excluded from $\mathcal{\tilde R}$ have multidegree $\{1,0\}$, $\{1,0\}$, and $\{0,1\}$, respectively. The corresponding Hilbert series is defined as 
\begin{align}
	H(\lambda,m) \equiv \sum\limits_{\alpha,\beta} (\dim \mathcal{R}_{\alpha\beta}) \lambda^\alpha m^\beta,
	\label{eq:HS_n_m}
\end{align}
with $\lambda$ and $m$ being dummy variables here. 

A closed expression for $H(t)$, $H(\lambda,m)$, and $H(q,y,t)$ can be constructed from pure group-theoretical considerations with obvious relations 
\begin{align}
	H(t) & = H(t,t), \qquad
	      H(\lambda, m)  =  \frac{1}{(1-\lambda)^2 (1-m)}H(\lambda,m,\lambda).  
	\label{eq:HS_grading_relations}
\end{align}

For Lie groups it can be computed in a number of ways. For example, one can  apply the Weyl integration formula or use the so-called Omega calculus (see, e.g., ref.~\cite{Bento:2021hyo} for details and application to multi-Higgs-doublet extensions of SM). In this paper, we also mention another way to compute $H(t)$ by the so-called \emph{free} resolution of the syzygy module (see appendix \ref{app:Hilbert_series}). 
The Hilbert series $H(t)$ was obtained for the first time in ref.~\cite{Bednyakov:2018cmx}
\begin{align}
	H(t) = \frac{1 + t^3 + 4 t^4 + 2 t^5 + 4 t^6 + t^7 + t^{10}}{
		\left(1-t\right)^3
		\left(1-t^2\right)^4
		\left(1-t^3\right)^3
		\left(1-t^4\right)
	},
	\label{eq:HS_ungraded}
\end{align}
while $H(q,y,t)$ was first presented in ref.~\cite{Trautner:2018ipq}.
The degree $d_i$ of $N=11$ \emph{algebraically independent} basis invariants $\{f_i\}$, $i=1,\ldots,N$ can be deduced from the factors $(1-t^{d_i})$ in the denominator of  $H(t)$.
As it was demonstrated in ref.~\cite{Trautner:2018ipq}, the set consists of 3 degree-one invariants, 4 degree-two invariants, 3 degree-three invariants and 1 degree-four invariant.

The number of invariants of a particular degree 
entering the generating set can be read off the \emph{first positive} 
terms in the expansion of the plethystic logarithm (see, e.g., ref.~\cite{Benvenuti:2006qr}) corresponding to $H(q,y,t)$ \cite{Trautner:2018ipq}. To save space, we provide the definition and expansion of plethystic logarithm constructed from $H(t)$: 
\begin{align}
	PL[ H(t) ] & \equiv \sum\limits_{k=1}^\infty  \frac{\mu(k) \ln H(t^k)}{k} = \sum\limits_{n=1}^\infty d_n t^{a_n} 
	\\  
		   &  = 3 t + 4 t^2 +4 t^3 +5 t^4 +2 t^5 + [4-1] t^6  \nonumber\\
		   & - 3 t^7 - 12 t^8 - 12 t^9 - [17 - 2] t^{10} - [8 - 14]t^{11} - [10 - 38] t^{12}  \ldots 
\label{eq:PL_unigraded}
\end{align}
where $\mu(k)$ is the M\"obius function. The coefficients in square brackets mix contributions with positive and negative signs, which can be explicitly separated in the multigraded case.   
One more important application of the plethystic logarithm is that the \emph{first negative} terms count the first syzygies and their corresponding degrees.
Indeed, summing the first ``positive'' coefficients, we obtain $M=22$, while the ``negative'' ones give $K=63$. In appendix~\ref{app:Hilbert_series} we provide more insights into the numbers appearing in \eqref{eq:PL_unigraded} by considering the free resolution of the syzygy module.

Given a generating set of the first syzygies, we can \emph{explicitly} construct \emph{linear-independent} invariants of $\mathcal{R}_{\alpha\beta}$ needed for the derivation of the RG equation for all the invariants from the generating set discussed in Sec.~\ref{sec:RG_algorithm}. 

\subsection{\label{sec:generating_set}Generating set of invariants}
In this section  we review the full generating set for the scalar sector of 2HDM constructed in ref.~\cite{Trautner:2018ipq}. The latter can be expressed in terms of $SO(3)$ covariants, given in Eqs.~\eqref{eq:lambda_to_Lambda}, \eqref{eq:m_to_M}. The invariants composed only from the self couplings can be cast into\footnote{In ref.~\cite{Trautner:2018ipq}, a different notation is used for the same quantities $T_{abc} \equiv \mathcal{I}_{abc}$ together with a different parameterization of the scalar potential. We express $\mathcal{I}_{abc}$ in terms of the Lagrangian parameters \eqref{eq:V_gen} and use simple linear algebra to obtain $\mathcal{I}_{abc}$ in terms of the $SO(3)$ covariants.  Let us mention that we use the definition for $\mathcal{I}_{300}$ from the main text of ref.~\cite{Trautner:2018ipq} not from the file attached to this reference in which the $\mathcal{I}_{300}$ invariant is given with a \emph{flipped sign}. ref.~\cite{Bento:2020jei} seems to utilizes the latter definition, i.e, $\mathcal{I}_{300} = -\T300$.}
\begin{align}
	4 \cdot Z_{1_{(1)}} &= 3 \Lambda_{00} + \trlam{}{}, \label{eq:Z11}\\
	4 \cdot Z_{1_{(2)}} &= \Lambda_{00} - \trlam{}{}, \label{eq:Z12}  \\
	8 \cdot \T200 &= \trlam{}{2}  - \frac{1}{3} \trlam{2}{} = \tr\tilde\lamT^2, \label{eq:T200}\\
		48 \cdot \T300& = \trlam{}{3} - \trlam{}{} \trlam{}{2} + \frac{2}{9} \trlam{3}{}= \tr\tilde\lamT^3, \label{eq:T300} \\
		4 \cdot \T002 & =   \lamV \cdot \lamV, \label{eq:T002}\\
	8 \cdot \T102& = \lamV \cdot \big[\lamT - \frac{1}{3} (\trlam{}{}) \big] \cdot \lamV
	= \lamV \cdot \tilde\lamT \cdot \lamV
, \label{eq:102}
	\\
	16 \cdot \T202 &    = 
 2 (\trlam{}{})\, (\lamV \cdot \lamT \cdot \lamV)  
 + \big[\trlam{}{2} - \frac{2}{3} \trlam{2}{}\big] \, (\lamV \cdot \lamV)
-3 \,(\lamV \cdot \lamT^2 \cdot \lamV) \nonumber\\
		       & = (\tr\tilde\lamT^2) \, (\lamV\cdot\lamV) - 3 (\lamV \cdot \tilde\lamT^2 \cdot \lamV)
. \label{eq:202}
\end{align}
The only CP-odd invariant constructed solely from the quartic couplings is 
\begin{align}
	64 i \cdot \T303 &= \lamV \cdot [(\lamT \cdot \lamV) \times (\lamT^2 \cdot \lamV)] 
= \lamV \cdot [(\tilde \lamT \cdot \lamV) \times (\tilde \lamT^2 \cdot \lamV)].
	\label{eq:303}
\end{align}
One also introduces invariants linear in the mass parameters. The CP-even set is given by
\begin{align}
	\Y1 &  = M_0, \label{eq:M0}\\ 
	4 \cdot \T011 & = \lamV\cdot\mV, \label{eq:T011}\\
	8 \cdot \T111 & = (\lamV\cdot\lamT \cdot  \mV) - \frac{1}{3} (\trlam{}{})\, (\lamV \cdot \mV)  = (\lamV \cdot \tilde \lamT \cdot \mV), \label{eq:T111}\\
	16 \cdot \T211 &  = 
			         2 (\trlam{}{})\, (\lamV \cdot \lamT \cdot \mV)
			      + \big[\trlam{}{2}
				       - \frac{2}{3}
				       \trlam{2}{}
			       \big]
				       (\lamV \cdot \mV)
			       - 3 (\lamV \cdot \lamT^2 \cdot \mV)
	\nonumber	       \\
		       & = (\tr \tilde \lamT^2)\,  (\lamV \cdot \mV)
			       - 3 (\lamV \cdot \tilde \lamT^2 \cdot \mV)
.\label{eq:T211}
\end{align}
while the CP-odd invariants can be represented as
\begin{align}
	16 i \cdot \T112 & = 	- \mV \cdot  [\lamV \times (\lamT \cdot \lamV) ] 
 = - \mV \cdot  [\lamV \times (\tilde \lamT \cdot \lamV) ] 
, \label{eq:T112}
 \\
	32 i \cdot \T212 & = 
	\mV  \cdot [\lamV \times (\lamT^2 \cdot \lamV)] 
	-\frac{2}{3} (\trlam{}{}) \mV \cdot [\lamV \times (\lamT \cdot \lamV) ] 
=	\mV  \cdot [\lamV \times (\tilde\lamT^2 \cdot \lamV)] 
, \label{eq:T212}
 \\
	64 i \cdot \T312 & = 
					 3 \mV \cdot [(\lamT\cdot \lamV) \times (\lamT^2\cdot \lamV)] 
					  - (\trlam{}{})\, \mV\cdot [\lamV\times(\lamT^2 \lamV)]
					  + (\trlam{}{2}) \, 
					  \mV  \cdot [\lamV \times (\lamT\cdot \lamV)]
	\nonumber			  \\
			 & = 
					 3 \mV \cdot [(\tilde \lamT\cdot \lamV) \times (\tilde \lamT^2\cdot \lamV)] 
					  + (\tr \tilde \lamT^2) \, 
					  \mV  \cdot [\lamV \times (\tilde \lamT\cdot \lamV)]
, \label{eq:T312}
\end{align}
In ref.~\cite{Trautner:2018ipq} the following CP-even invariants quadratic in mass are introduced
\begin{align}
	4 \cdot \T020 & = \mV\cdot\mV, \label{eq:T020}\\
	8 \cdot \T120 & = (\mV\cdot\lamT \cdot  \mV) - \frac{1}{3} (\trlam{}{})\, (\mV \cdot \mV)  = (\mV \cdot \tilde \lamT \cdot \mV), \label{eq:T120}\\
	16 \cdot \T220 &  = 
			         2 (\trlam{}{})\, (\mV \cdot \lamT \cdot \mV)
			      + \big[\trlam{}{2}
				       - \frac{2}{3}
				       \trlam{2}{}
			       \big]
				       (\mV \cdot \mV)
			       - 3 (\mV \cdot \lamT^2 \cdot \mV)
	\nonumber	       \\
		       & = (\tr \tilde \lamT^2)\,  (\mV \cdot \mV)
			       - 3 (\mV \cdot \tilde \lamT^2 \cdot \mV)
. \label{eq:T220}
\end{align}
One can also introduce CP-odd invariants quartic in mass  
\begin{align}
16 i \cdot \T121 & = 	- \lamV \cdot  [\mV \times (\lamT \cdot \mV) ] 
 = - \lamV \cdot  [\mV \times (\tilde \lamT \cdot \mV) ] 
, \label{eq:T121}
 \\
	32 i \cdot \T221 & = 
	\lamV  \cdot [\mV \times (\lamT^2 \cdot \mV)] 
	-\frac{2}{3} (\trlam{}{}) \lamV \cdot [\mV \times (\lamT \cdot \mV) ] 
=	\lamV  \cdot [\mV \times (\tilde\lamT^2 \cdot \mV)] 
, \label{eq:T221}
 \\
	64 i \cdot \T321 & = 
					 3 \lamV \cdot [(\lamT\cdot \mV) \times (\lamT^2\cdot \mV)] 
					  - (\trlam{}{})\, \lamV\cdot [\mV\times(\lamT^2 \mV)]
					  + (\trlam{}{2}) \, 
					  \lamV  \cdot [\mV \times (\lamT\cdot \mV)]
	\nonumber			  \\
			 & = 
					 3 \lamV \cdot [(\tilde \lamT\cdot \mV) \times (\tilde \lamT^2\cdot \mV)] 
					  + (\tr \tilde \lamT^2) \, 
					  \lamV  \cdot [\mV \times (\tilde \lamT\cdot \mV)]
. \label{eq:T321}
\end{align}
Finally, a CP-odd invariant trilinear in masses \cite{Trautner:2018ipq} is given by 
\begin{align}
	64 i\cdot \T330  & =  \mV \cdot [(\lamT \cdot \mV) \times (\lamT^2 \cdot \mV)] 
= \mV \cdot [(\tilde \lamT \cdot \mV) \times (\tilde \lamT^2 \cdot \mV)].
\label{eq:T330}
\end{align}
In ref.~\cite{Trautner:2018ipq,Bento:2020jei}, the following $N=11$ invariants are chosen to be algebraically independent
\begin{align}
	\underbrace{\Z11, \Z12, \Y1}_{\text{degree 1}}, 
	\underbrace{\T200, \T002, \T020, \T011}_{\text{degree 2}}, 
	\underbrace{\T300, \T102, \T120}_{\text{degree 3}}, 
	\underbrace{\T211}_{\text{degree 4}}. 
\label{eq:alg_indep_invs_unigraded} 
\end{align}
In appendix.~\ref{app:algebraic_dependence} we provide the relations of the form \eqref{eq:algebric_relation} between the remaining invariants from the generating set and the set \eqref{eq:alg_indep_invs_unigraded}. 

Note also that the CP-odd invariants, highlighted in red, can only appear in expressions linearly. This fact can easily be deduced, for example,  from geometrical considerations. It turns out that the same is true for the two CP-even invariants $\T111$, and $\T211$, highlighted in blue. In the next section we demonstrate that, due to syzygies, high powers and mixed products of $\T111$ and $\T211$  can be reduced to expressions linear in $\T111$ and $\T211$.    

\subsection{\label{sec:syzigies}First syzygies}

As mentioned earlier, we have $K=63$ first syzygies that generate $\IdealTHDM$.  In principle, they can be found systematically by the linear-algebra method proposed in ref.~\cite{Trautner:2018ipq}, i.e., one can consider a $\mathtt{k}$-linear combination of all possible power products of invariants $g_n$ (monomials $m^{(i)}_{abc}$) at a certain degree $[abc]$:
\begin{align}
	\sum\limits_{i=0}^{n_0} c_{abc}^{(i)} m^{(i)}_{abc} \stackrel{?}{=}0, 
\end{align}
express the latter, e.g., in terms of the couplings and masses, and equate to zero the coefficients of monomials constructed from the Lagrangian parameters. 
An alternative but still brute-forced approach  was used in ref.~\cite{Wang:2021wdq}: a system of linear equations on $n_0$ coefficients ($c_{abc}^{(i)}$ in our case) can be obtained by expressing $g_n$ entering $m^{(i)}_{abc}$ in terms of physical observables and sampling the latter with random values $n_0$ times.  
With this kind of  techniques, we can derive all possible relations at a certain (multi) degree. However, for higher degrees it becomes harder and harder to decide whether a relation is just a consequence of lower-degree ones, or indeed a new one (see discussion in appendix E. of ref.~\cite{Trautner:2018ipq}). Speaking differently, it is not a straightforward task to decide if a relation is a new generator of the full ideal $\IdealTHDM$, or it belongs to the ideal generated by the lower-degree generators. 

To overcome this difficulty, we have used the Gr\"obner-basis techniques (see appendix \ref{app:gb_syzygies}) implemented in the \Macaulay\ software\footnote{One can also use other packages for similar problems, e.g., \texttt{Singular} \cite{DGPS}.} to explicitly derive the needed relations that generate $\IdealTHDM$ for a deliberately chosen monomial ordering in the polynomial ring $\mathcal{R}$.  After computing minimal generators $\{r_k\}$ of $\IdealTHDM$, we solve $r_k = 0$ for the leading term, $\LT(r_k)$, entering $r_k$, and present the resulting expressions below. 

Let us mention that 3 out of 63 computed syzygies allow one to eliminate the products $\T111^2$, $\T211^2$, and $\T211 \T111$ of two CP-even invariants  (``even$\times$even'' syzygies)
\begin{align}
	{}^6[222]_{\phantom{1}}: && 3 \T111^2 & = 2 \T011 \T211+3 \T102 \T120 + (\T002 \T020 - \T011^2) \T200    
	\nonumber\\
	      &&&
- \T020 \T202  - \T002 \T220, 
\label{eq:syz_222}\\ 
{}^7[322]_{\phantom{1}}: && 2 \T111 \T211 & = 4 \T011 \T200 \T111  -2 \T200 (\T020 \T102 +2 \T002 \T120 ) \nonumber \\
      &&  & + \T120 \T202 + \T102 \T220 - 6 \T300 (  \T002 \T020 - \T011^2 ) 
\label{eq:syz_322}\\ 
{}^8[422]_{\phantom{1}}: && \T211^2 & = - 2 \T011 (18 \T300 \T111 +  \T200 \T211) 
	- (\T002 \T020  - \T011^2) \T200^2  \nonumber \\
      &&	& + \T200 (\T020  \T202 + \T002 \T220 )+ \T202 \T220\nonumber\\
      &&&
	+ 18 \T300 (\T020 \T102 + \T002 \T120 ), 
\label{eq:syz_422} 
\end{align}
in favour of expressions linear in $\T111$, and $\T211$. For convenience, we use the notation ${}^x[abc]_i$ to specify the total degree $x$ of the $i$-th relation corresponding to a multidegree $[abc]$. The subscript $i$ is absent if there is a single $[abc]$-syzygy. 
The remaining 60 relations can be split into  24 ``odd $\times$even'' syzygies (note that an $[acb]$-syzygy can be obtained from the $[abc]$ one by a simple index permutation)
{\allowdisplaybreaks
\begin{align}
{}^7[223]_{\phantom{1}}: && \T111 \T112 & = \T011 \T212 + \T002 \T221- \T102 \T121, 
\label{eq:syz_223} \\ 
{}^8[323]_1: && 3 \T111 \T212 & = \T200 (\T011 \T112 + \T002 \T121) \nonumber \\
	     &&& - \T202 \T121  - \T011 \T312 + \T002 \T321, \label{eq:syz_323_1_of_3}\\ 
{}^8[323]_2: && \T211\T112  & = \T200(\T011 \T112 + \T002 \T121 ) \nonumber \\
	     &&& - 3 \T102 \T221 + \T011 \T312 - \T002 \T321, \label{eq:syz_323_2_of_3} \\
{}^8[323]_3: && 3 \T020 \T303 & = \T202 \T121 - 3 \T102 \T221 + 3 \T011 \T312 - 2 \T002 \T321, 
\label{eq:syz_323_3_of_3}\\ 
{}^{8}[314]_{\phantom{1}}: && 3 \T011 \T303 & = \T202 \T112 - 3 \T102 \T212 + \T002 \T312, 
\label{eq:syz_314}\\ 
{}^9[423]_1: && \T111 \T312 & =  \T200 (\T102 \T121 - 2 \T011  \T212) - \T202 \T221 \nonumber\\
       &&&- 6 \T300 \T011 \T112  + \T102 \T321, 
\label{eq:syz_423_1_of_3}\\ 
{}^9[423]_2: && \T211 \T212 & = -\T200 (\T011 \T212 + \T002 \T221) - \T202 \T221 \nonumber\\
       &&& - 6 \T300 (\T011 \T112  + \T002 \T121), \label{eq:syz_423_2_of_3}\\ 
{}^9[423]_3: && 3 \T120 \T303 & = 2 (\T200 \T102 - 3 \T300 \T002 )\T121 - 18 \T300 \T011 \T112 - 6 \T011 \T200 \T212 \nonumber \\
       &&	      & - 2 (\T200 \T002  + \T202) \T221 + \T102 \T321, 
\label{eq:syz_423_3_of_3}\\ 
{}^9[414]_{\phantom{1}}: && 	3 \T111 \T303 & = -\T200 (\T102 \T112  + 2 \T002  \T212) + \T202 \T212 \nonumber \\
       &&& - 6 \T300 \T002 \T112 + \T102 \T312, \label{eq:syz_414}\\ 
       {}^{10}[523]_1: && \T211 \T312 & = \T200 (2 \T200 \T011  \T112 - \T202 \T121 - 3 \T102 \T221) \nonumber\\
       &&& - 18 \T300 (\T102 \T121  - \T011 \T212) + \T202 \T321,
\label{eq:syz_523_1_of_2}\\ 
%
{}^{10}[523]_2:&&3 \T220 \T303 & = 6 \T200^2 \T011  \T112 + 2 \T200 (\T002 \T200-   \T202) \T121  
- 36 \T300 \T102  \T121  
\nonumber \\
      &&	      & + 54 \T300 \T011 \T212  - 6 (\T200 \T102  - 3 \T300\T002 ) \T221 + \T202 \T321,\label{eq:syz_523_2_of_2}\\
{}^{10}[514]_{\phantom{1}}:&&3 \T211 \T303 & = \T200 \left[(2 \T200 \T002  +  \T202) \T112 + 3  \T102 \T212\right]
\nonumber \\
      &&      & 
+ 18 \T300 (\T102 \T112  + \T002 \T212)  + \T202 \T312,
\label{eq:syz_514} 
\end{align}
}
and 36 ``odd $\times$odd'' syzygies 
allowing to express the product of any 2 out of 8 odd invariants from the generating set  in terms of even invariants: 
{\allowdisplaybreaks
\begin{align}
{}^{8}[233]_{\phantom{1}}: && 3 & \T112  \T121 = (3 \T011^2 - \T002 \T020) \T211 -3 (\T020 \T102  + \T002 \T120) \T111 \nonumber \\
			   &&& + 3 \T200  \T011 (\T002\T020 - \T011^2) 
			   + 6 \T011 \T102 \T120 
			   \nonumber \\
      &&    & - \T011 (\T020 \T202  + \T002 \T220), \label{eq:syz_233}\\ 
	{}^8[224]_{\phantom{1}}: && 3 &\T112^2 = 2 \T002 \T011 \T211  - 6 \T011 \T102 \T111 + 3 \T020 \T102^2  + 3 \T002 \T102 \T120 \nonumber \\
	       && &- \T200 \T002 (\T002 \T020 -  \T011^2  ) - \T011^2 \T202 - \T002^2 \T220, 
\label{eq:syz_224}\\ 
{}^{9}[324]_{\phantom{1}}: && 6 & \T112  \T212  =  2 \T011 ( 2\T200 \T002  - \T202 ) \T111- 2 \T011 \T102 \T211 
+ \T002 \T102 \T220 \nonumber \\
       &&     & -2 \T200 (\T011^2 \T102 +\T002^2 \T120)  + (2 \T020 \T102 + \T002 \T120) \T202 \nonumber \\
       &&      & - 6 \T300 \T002( \T011^2 -  \T002 \T020 ), 
\label{eq:syz_324}\\ 
{}^{9}[333]_1: && 6 & \T121  \T212 = 2 \left[ \T200 (3 \T011^2 -   \T002 \T020 ) - \T020 \T202 \right]\T111 - 2 \T002 \T120 \T211
\nonumber \\
       &&     & -2 \T200 \T011 (\T020 \T102  +  \T002  \T120)+ \T011 (3 \T120 \T202 + \T102 \T220) \nonumber \\
       &&      &   + 18 \T300 \T011 (\T011^2  - \T002 \T020),
\label{eq:syz_333_1_of_2}\\ 
	{}^{10}[424]_1:&& 9 & \T212^2  =  - 36 \T300 \T011 \T002 \T111 
		- 2 \T011 (\T002 \T200 + \T202) \T211 \nonumber\\
	      && & + \T200^2 \T002(\T011^2  - \T002 \T020) 
		+ \T200 (\T011^2 \T202 + \T002^2 \T220 ) \nonumber\\
	      && & + (\T020 \T202 + \T002 \T220 ) \T202 
		+ 18 \T300 (\T011^2 \T102 + \T002^2 \T120),
\label{eq:syz_424_1_of_3}\\ 
{}^{10}[424]_2: && 6  & \T112 \T312  = 6 \T011 (\T200 \T102  - 6 \T300 \T002 )\T111 
- 2 \T011 (2 \T002 \T200 + \T202 ) \T211\nonumber \\
       &&     &
	      - 2 \T200 \left[3  \T002 \T102 \T120 -  (\T011^2 +  \T002 \T020) \T202\right]
+ 2 \T002 \T202 \T220 
	      \nonumber\\
       &&      & - 3 \T102 (\T120 \T202 - \T102 \T220) 
	      + 18 \T300 (\T011^2 + \T002 \T020 ) \T102,
      \label{eq:syz_424_2_of_3}\\ 
{}^{10}[424]_3: && 18  & \T121 \T303 
 = 
	  18 \T011 (6 \T300 \T002  - \T200 \T102) \T111  
	+ 6  \T011 (2 \T200 \T002  + \T202) \T211  \nonumber\\
&&&
	+ 6 \T200 (\T020 \T102  + 2 \T002  \T120 ) \T102 
	+ 3 (\T102 \T120 - 2 \T200 \T011^2 )\T202  \nonumber\\
&&&
	- 2 \T200 (\T020 \T202 + 2  \T002 \T220 )\T002	
	- 2 (\T020 \T202 + 2 \T002 \T220) \T202 
	\nonumber\\
&&&
	- 3 \T102^2 \T220 
	- 18 \T300 (\T102(3  \T011^2 +  \T002 \T020) + 2 \T002^2 \T120), 
\label{eq:syz_424_3_of_3}\\ 
{}^{10}[433]_1: && 3  & \T121\T312  = 3 (\T120 \T202 -\T200 \T020 \T102  + 18 \T300 \T011^2 )\T111 
\nonumber \\
	       &&  &
+ ( 6 \T200 \T011^2 - 3 \T102 \T120  + \T020 \T202 ) \T211  \nonumber \\
	       && &
+ \T200  \T011 (3 \T102 \T120 - 4 \T020 \T202 - 2 \T002 \T220)   \nonumber \\
	       && &- \T011 [\T202 \T220 + 18 \T300 (2 \T020 \T102 +  \T002 \T120)],
\label{eq:syz_433_1_of_3} \\
{}^{10}[433]_3: && 9  & \T212 \T221 
        = 
	(\T002 \T020  - 3 \T011^2) (18 \T300 \T111  + \T200 \T211)
	\nonumber\\
		&&&
	- (\T020 \T202 + \T002 \T220 )\T211  
       + 2 \T011 \T202 \T220 
	\nonumber\\
       && & 	
       - 3 \T200^2 \T011 (\T011^2 -  \T002 \T020)
	+ \T200 \T011 ( \T020 \T202 + \T002 \T220)  \nonumber\\
       && & 	
	+ 18 \T300 \T011 (\T020 \T102 + \T002 \T120),
\label{eq:syz_433_3_of_3} \\
{}^{10}[415]_{\phantom{1}}: && 9 & \T112 \T303 
	        = 
- 3 (\T200 \T002 \T102 + 6 \T300\T002^2 +  \T102 \T202 )\T111  
\nonumber
\\
	       &&&
+ (\T002 \T202 - 2 \T200 \T002^2  + 3 \T102^2) \T211 
	\nonumber\\
	       &&& 
+ \T200 \T011 (3 \T102^2  + 2 \T002 \T202) 
	- \T011 (\T202^2 - 18 \T300 \T002 \T102),
\label{eq:syz_415}
\end{align}
\begin{align}
	{}^{11}[524]_1: && 6  & \T212 \T312 
 = 
2 \T011 [ \T200 (2 \T200 \T002 -  \T202)  - 18 \T300 \T102] \T111  \nonumber\\
&&&
	- 2 \T011 (\T200 \T102 - 6 \T300 \T002 ) \T211 
	 -2 \T200^2 (\T011^2  + \T002 \T020 ) \T102 \nonumber\\
&&&
	+ 2 \T200  (\T120 \T202 +  \T102 \T220) \T002  
	+ \T202 (\T102 \T220 - \T120 \T202) \nonumber \\
&&&
	+ 6 \T300 [6 \T002 \T102 \T120 -  (\T011^2 +  \T002 \T020) \T202],
\label{eq:syz_524_1_of_2}
	\\ 
{}^{11}[524]_2: && 18  & \T221 \T303 = 
6 \T011 [  \T200 (\T202 -2 \T002 \T200) + 18 \T300 \T102 ]\T111  
\nonumber\\
       &&     &
	+ 6 \T011 (\T200 \T102 - 6 \T300 \T002 )\T211  
	+  (\T120 \T202 - \T102 \T220) \T202 
\nonumber\\
	       &&      &
	+ 6 \T300 [3 \T011^2 \T202 
	+ \T020 (\T002 \T202-6 \T102^2 ) + 2 \T002 (\T002 \T220-6 \T102 \T120)]   
\nonumber\\
	       &&&
+ \T200^2 [6 \T011^2 \T102 + 2 \T002 (\T020 \T102 + 2 \T002 \T120)] 
	\nonumber\\
	       &&&
	- 4 \T200 [\T020 \T102 \T202 + \T002 (\T120 \T202 + \T102 \T220)] 
\label{eq:syz_524_2_of_2}
\\
{}^{11}[533]_1: && 3  & \T221 \T312 
 = 
	6(3 \T300 \T020 \T102  -  \T200^2 \T011^2 ) \T111 
	+ \T202 (\T200 \T020   + \T220 ) \T111  \nonumber\\
&&&
	+ [\T102 (\T020 \T200 -  \T220) - 18 \T300 \T011^2]  \T211  
	- \T200 (\T120  \T202 + \T102 \T220)\T011 
	\nonumber\\
&&&
	+ 6 \T300 (\T002 \T220 + 2 \T020 \T202 - 3 \T102 \T120 ) \T011 \nonumber\\
&&&
+ 2 \T200^2 (2 \T020 \T102  +  \T002  \T120) \T011 
\label{eq:syz_533_1_of_2}
	\\ 
{}^{11}[515]_{\phantom{1}}:&& 9  & \T212 \T303 
	 = 
	 [\T200 (2 \T200 \T002^2  + \T002 \T202) - \T202^2 + 18 \T002 \T102 \T300] \T111 
	\nonumber\\
	&&	&
	+ [\T002(\T200 \T102 + 6 \T300 \T002) + \T102 \T202 ] \T211  
	\nonumber\\
	&&&
	   -2 \T200 (\T200 \T002 + \T202)\T011 \T102  
	- 6  \T300 (3 \T102^2 + \T002 \T202)\T011 
\label{eq:syz_515}
\\
{}^{12}[633]_1: && 27  & \T303 \T330  = 
3 [ 
	\T200 (\T120 \T202 + \T102 \T220) 
	+ 18  \T300  
	         \T102 \T120 ]\T111 
		 \nonumber \\
		 &&& + 	       [
		3 \T200 \T102 \T120 -\T202 \T220 
		] \T211 
\nonumber\\
		 &&&
	-6 (\T200^2 \T111 + 3 \T300 \T211)(\T020 \T102 + \T002 \T120)
	\nonumber\\
		 &&&
	-2 (\T200 \T211 + 9 \T300 \T111) (\T020 \T202 + \T002 \T220)
\nonumber \\
		 &&& 
		+ 7 \T200  (\T202 \T220 + 3 \T200 \T102 \T120)  \T011  
		\nonumber\\
		 &&&
	+ 2  ( 7 \T002 \T020 -9 \T011^2) ( \T200^3  - 27 \T300^2)\T011  
	\nonumber \\
		 &&&
	+ 63 \T300 (\T120 \T202 + \T102 \T220) \T011  
\label{eq:syz_633_1_of_2}
\\
	{}^{12}[633]_2: && 3  & \T312 \T321  = 
3 [\T200 (\T120 \T202 + \T102 \T220) 
	+ 18 \T300 \T102 \T120 ]\T111  
\nonumber
	\\
		 && &
+ \T211 (3 \T200 \T102 \T120  - \T202 \T220) 
+ 27 \T300 (\T120 \T202 + \T102 \T220) \T011
\nonumber\\
		 && &
+ 2 (\T200^3 - 27 \T300^2) (\T002 \T020-3 \T011^2) \T011
\nonumber\\
		 && &
+ 3 \T200 (\T202 \T220  + 3 \T200 \T102 \T120) \T011,
\label{eq:syz_633_2_of_2}
\\
	{}^{12}[615]_{\phantom{1}}: && 9  & \T312\T303  = 
6 [\T200 (\T200 \T002 +  \T202) \T102
	+ 3 \T300 (\T002 \T202 + 3 \T102^2 )] \T111  
	\nonumber \\
       &&      &
+ [\T200 (2 \T002 \T202 + 3 \T102^2) + 18 \T300 \T002 \T102  -\T202^2 ]\T211  
\label{eq:syz_615}
\\
       &&      &
  + [\T200   \T202^2 
 + 3 (\T200^2 \T102
+ 6 \T300   \T202)\T102 - 4  (\T200^3 - 27 \T300^2) \T002^2 
] \T011,
\nonumber \\
{}^{12}[624]_1:&& 3 & \T312^2 
 = 
	12 \T011 (\T200^2 \T102 + 3  \T300 \T202)\T111   
	+ 4 \T011 (\T200 \T202 + 9 \T300 \T102 ) \T211
	\nonumber\\
& &&
	+ 4(27 \T300^2 - \T200^3 )  \T011^2 \T002 
	+ 3 [\T200 (\T220 - \T200 \T020) 
	+ 18 \T300 \T120] \T102^2 
	\nonumber\\
&& &
	+ 6 (\T200 \T120 - 3 \T300 \T020) \T102 \T202  
	-  (\T200 \T020 + \T220) \T202^2,
\label{eq:syz_624_1_of_2}\\
{}^{12}[624]_2: && 9  & \T321 \T303 = 6 \T011[3 (\T200^2 \T102 + 3 \T300 \T202 )\T111  
+ (\T200 \T202 + 9 \T300 \T102 )\T211  
]
\nonumber \\
	&& &
+ \T200 [ 3 (\T102 \T220 +  2 \T120 \T202)\T102  - (\T002 \T220 + 2 \T020 \T202 )\T202 ] 
- \T202^2 \T220 
\nonumber\\
	&&     &
-3 \T200^2 (\T002 \T120 + 2 \T020 \T102 ) \T102  
+ 2  (\T200^3 - 27 \T300^2) (\T002 \T020 -3 \T011^2) \T002  
\nonumber\\
	&&     &
- 9 \T300 [ \T002 (\T120 \T202 + \T102 \T220) + 2 (2 \T020 \T202- 3 \T102 \T120) \T102  ],
\label{eq:syz_624_2_of_2}\\
{}^{12}[606]_{\phantom{1}}: && 27 &\T303^2  = 
	  3 \T200 (
	    \T002 \T202^2 
	+ 3 \T102^2 \T202 
	+ 3 \T200 \T002 \T102^2) \nonumber \\ 
       &&   &
	- 4 (\T200^3  - 27 \T300^2)\T002^3 
	+ 54 \T300 (\T102^2 + \T002 \T202 )\T102
	- \T202^3 
	.
\label{eq:syz_606}
\end{align}
}
We discuss the application of syzygies to the derivation of six-loop RG functions for the basis invariants in the next section. 

\section{\label{sec:RG}Renormalization group equations for basis invariants}
\subsection{\label{sec:RG_algorithm}Derivation of RG: an algorithm}

As mentioned earlier, we cannot associate a particular triplet $[abc]$ or a doublet $\{\alpha,\beta\} $ to a multi-loop RG function for an invariant. Each new loop adds powers of the self-couplings but the mass dimension of the invariant does not change. 
Nevertheless, with each loop contribution, one can associate a particular multidegree\footnote{This is only true if we neglect all other couplings of 2HDM. For example, gauge couplings, which are singlets w.r.t the considered global symmetry, will definitely prevent us to do this, thus, invalidating the anzatz \eqref{eq:RGE_gen_form_invs}. Nevertheless, the latter can be extended to account for other couplings. We thank the anonymous referee for pointing out this subtlety.} $\{\alpha,\beta\}$, so that we can write schematically the RG equation for an invariant $I_{\alpha,\beta}$ from the generating set $\{g_n\}$ as 
\begin{align}
	\frac{d}{d t} I_{\alpha,\beta} = \sum\limits_{l=1}^{L} h^l \cdot \sum\limits_{i=1}^{d_{\alpha+l,\beta}} c^{(i)}_{\alpha,\beta;l} I^{(i)}_{\alpha+l,\beta},   \qquad t = \ln \mu
	\label{eq:RGE_gen_form_invs}
\end{align}
where $h = 1/(16 \pi^2)$ counts loops,  $I^{(i)}_{\alpha+l,\beta}$ $i = 1\ldots d_{\alpha+l,\beta}$ are $\mathtt{k}$-\emph{linear-independent} invariants (i.e., a ``basis'') of degree $\{\alpha+l,\beta\}$
and $c^{(i)}_{\alpha,\beta;l} \in \mathtt{k}$ are the corresponding numerical coefficients. The number $d_{\alpha,\beta}$ of linear-independent invariants with the degree $\{\alpha,\beta\}$ can easily be found from the multigraded Hilbert series \eqref{eq:HS_n_m} that we give here in the form:

\begin{align}
	H(\lambda,m)  & = \frac{N(\lambda,m) \cdot (1-\lambda)^{-2} \cdot (1-m)^{-1}}
	{
	\left(1-\lambda ^2\right)^2 
	\left(1-\lambda ^3\right)^2
	\left(1-\lambda ^4\right)
	\left(1- \lambda m \right) 
	\left(1-m^2\right) 
	\left(1- \lambda m^2\right)
	\left(1-\lambda ^2 m^2 \right)
}
\label{eq:HS_n_m_grading}
\end{align}
The denominator corresponds to the invariants
\begin{align}
	\underbrace{\Z11,\Z12}_{\lambda}, 
	\underbrace{\vphantom{\Z11}\Y1}_{m}, 
	\underbrace{\vphantom{\Z11} \T200, \T002}_{\lambda^2}, 
	\underbrace{\vphantom{\Z11} \T300, \T102}_{\lambda^3}, 
	\underbrace{\vphantom{\Z11} \T202}_{\lambda^4}, 
	\underbrace{\vphantom{\Z11} \T011}_{\lambda m}, 
	\underbrace{\vphantom{\Z11} \T020}_{m^2}, 
	\underbrace{\vphantom{\Z11} \T120}_{\lambda m^2}, 
	\underbrace{\vphantom{\Z11} \T220}_{\lambda^2 m^2}, 
\label{eq:hs_den_m_n} 
\end{align}
while the numerator looks like
\begin{align}
	N(\lambda,m)  = 
	1 + & \underbrace{m \lambda^2 (1 + \lambda) - m^3 \lambda^7(1 + \lambda) - m^4 \lambda^{10}}_{even} \\
	  + & \underbrace{\lambda^6 + m ( \lambda^3 + \lambda^4 + \lambda^5 - \lambda^7)
	  + m^2 ( \lambda^2 + \lambda^3 + \lambda^4 - \lambda^6 - \lambda^7 - \lambda^8)}_{odd} \nonumber \\
		  + & \underbrace{m^3 ( \lambda^3 - \lambda^5 - \lambda^6 - \lambda^7)
		  - m^4 \lambda^4}_{odd}.
\label{eq:HS_n_m_grading_numerator}
\end{align}
Here we indicate the terms corresponding to even and odd invariants. While the $\{\alpha,\beta\}$ grading does not immediately allow us  to deduce 
whether an invariant $I_{\alpha,\beta}$ is even or odd, we can trace this property from $H(q,y,t)$ \eqref{eq:HS_grading_relations}.  

In Eq.~\eqref{eq:HS_n_m_grading_numerator} the terms with positive coefficients in the numerator can be associated with the even invariants $\T111$ (multidegree $\{2,1\}$) and $\T211$ (multidegree $\{3,1\}$), 
and the odd invariants $\T303$ ($\lambda^6$), 
$\T112$ $(\lambda^3 m)$, $\T212$ ($\lambda^4 m$), $\T312$ ($\lambda^5 m$), 
$\T212$ $(\lambda^2 m^2)$, $\T221$ ($\lambda^3 m^2$), $\T321$ ($\lambda^4 m^2$), and
$\T330$ ($\lambda^3 m^3)$.  
By putting these terms in the numerator with a positive sign, we effectively avoid counting the corresponding powers and mixed products as independent invariants. In such a way we exclude  $\T111^2$, $\T111 \T303$, etc, thus, partially accounting for the syzygies. However, by expanding the denominator of \eqref{eq:HS_n_m_grading}, we can still obtain terms corresponding to
eight products $\T011\T303$ ($m \lambda^7$),  $\T020\T303$ ($m^2 \lambda^6$), $\T120 \T303$ ($m^2\lambda^7$), $\T220\T303$ ($m^2\lambda^8)$, $\T002\T330$  ($m^3\lambda^5$), $\T102\T330$ ($m^3 \lambda^6$), $\T202\T330$ ($m^3\lambda^7$), and $\T011\T330$ ($m^4\lambda^4$),  which we know can be reduced via the syzygies\footnote{Together with the ``permuted'' ones.} \eqref{eq:syz_323_3_of_3},\eqref{eq:syz_314},\eqref{eq:syz_423_3_of_3},\eqref{eq:syz_523_2_of_2}. This overcounting for odd invariants is fixed by the corresponding negative terms in the numerator \eqref{eq:HS_n_m_grading_numerator}. 

The situation is a bit more subtle for the even negative terms present in $N(\lambda,m)$. They reflect the fact that there are  non-trivial relations, linear in $\T111$ and $\T211$, and corresponding to higher syzygies. These relations can be obtained by considering two possible ways to reduce the products $\T111^2 \T211$ ($m^3 \lambda^7$), $\T211^2 \T111$ ($m^3 \lambda^8$), and $\T111^2 \T211^2$ ($m^4\lambda^{10}$) via the syzygies \eqref{eq:syz_222}, \eqref{eq:syz_322}, and \eqref{eq:syz_422}.

\begin{table}[t]
	\centering
\begin{tabular}{rccccccccccc}
     & $Z_{1_{(1)}}$ & $\T200$ & $\T300$ & &   &   &   &   &    &    &    \\
     & $Z_{1_{(2)}}$ & $\T002$ & $\T102$ & $\T202$ &   &   &   &   &  &   &    \\
 \hline
 $\alpha$ & 2 & 3 & 4 & 5 & 6 & 7 & 8 & 9 & 10 & 11 & 12 \\
 \hline
 {even} & \bf{5} & \bf{10} & \bf{19} & \bf{32} & \bf{54} & \bf{84} & \bf{129} & \bf{190} & \bf{275} & 386 & 536 \\
 {odd} & 0 & 0 & 0 & 0 & 1 & \bf{2} & \bf{5} & \bf{10} & \bf{19} & \bf{32} & \bf{54} \\
 \hline
        &   &   &  &  &  & $\T303$ &   &   &   &   &   \\
\end{tabular}

\caption{The number $d^{\text{even}}_{\alpha,0}$ ($d^{\text{odd}}_{\alpha,0}$) of CP-even (CP-odd) linear-independent invariants with degree $\{\alpha,0\}$. The one-loop RG equations for the invariants that appear in the column $\alpha$ are linear combinations of $d_{\alpha,0}$ terms. For example, the one-loop RGE for $\T303$ (see Eq.~\eqref{eq:b1_T303}) is a combination of $2$ terms: $Z_{1_{(1)}} \T303$ and $Z_{1_{(1)}} \T303$ corresponding to $d^{\text{odd}}_{\alpha,0}$. In bold we indicate all the cases that require explicit construction of all $I^{(i)}_{\alpha,0}$, e.g., six-loop RGE for $\T202$ involves $d^{even}_{10,0}=275$ linear-independent ring elements, while $\T303$ requires $d^{odd}_{12,0}=54$ elements at six loops.} 
\label{tab:d_n_0}
\end{table}

Equation \eqref{eq:RGE_gen_form_invs} suggests that if we know how to express all the invariants entering into the equation  in terms of couplings and masses, given  the $l$-loop contribution to the beta functions $\mathbf{\beta}^{(l)}_{\lambda_i}$ and $\beta^{(l)}_{m^2_{ij}}$,  
the RG equation for $I_{\alpha,\beta}$ can be found by  linear-algebra methods. Indeed, equating the coefficients of monomials built from different powers of $\lambda_i$ and $m^2_{ij}$ entering into
\begin{align}
	\left[\beta^{(l)}_{\lambda_i} \partial_{\lambda_i} + 
	\beta^{(l)}_{m^2_{ij}} \partial_{m_{ij}^2}  \right] I_{\alpha,\beta}(\lambda_k, m^2_{pq}) =  
	\sum\limits_{i=1}^{d_{\alpha+l,\beta}} c^{(i)}_{\alpha,\beta;l} \cdot I^{(i)}_{\alpha+l,\beta} (\lambda_{k},m^2_{pq}),
\end{align}
we obtain a system of linear equations on $c_{\alpha,\beta;l}^{(i)}$, which can be solved by \Mathematica \cite{Mathematica}, or, for higher loops, by means of \fermat \cite{fermat}.
We provide the numbers $d_{\alpha,\beta}= d^{\text{even}}_{\alpha,\beta} + d^{\text{odd}}_{\alpha,\beta}$ relevant for six-loop computations and corresponding to the CP-even and CP-odd invariants of the degree $\{\alpha,\beta\}$ 
in tables~\ref{tab:d_n_0} ($\beta = 0$), 
\ref{tab:d_n_1} ($\beta=1$), \ref{tab:d_n_2} ($\beta=2$), and \ref{tab:d_n_3} ($\beta=3$).     

Let us stress once again that the syzygies play a major role in the explicit construction of $I^{(i)}_{\alpha,\beta}$ since naive products of 
$g_n$ can overcount $d_{\alpha,\beta}$.
It is relatively easy to generate all such monomials at a sufficiently high degree, e.g., by a computer. 
To construct a basis, we have to decide which of these products (or, in general, combinations) have to be removed to get the correct number $d_{\alpha,\beta}$ of basis elements.

\begin{table}[t]
	\centering
\begin{tabular}{rccccccccccc}
      & $Y$ & $\T011$ & $\T111$ & $\T211$ &   &   &   &   &   &   &   \\
	\hline
{$\alpha$} & 1 & 2 & 3 & 4 & 5 & 6 & 7 & 8 & 9 & 10 & 11 \\
	\hline
{even} & \bf{3} & \bf{8} & \bf{18} & \bf{36} & \bf{66} & \bf{115} & \bf{189} & \bf{299} & \bf{457} & 678 & 980 \\
 {odd} & 0 & 0 & 1 & \bf{3} & \bf{8} & \bf{18} & \bf{36} & \bf{66} & \bf{115} & \bf{189} & \bf{299} \\
	\hline
      &   &  &  & $\T112$ & $\T212$  & $\T312$   &   &   &   &   &   \\
\end{tabular}
\caption{The number $d^{\text{even}}_{\alpha,1}$ ($d^{\text{odd}}_{\alpha,1}$) of CP-even (CP-odd) linear-independent invariants with multidegree $\{\alpha,1\}$. } 
\label{tab:d_n_1}
\end{table}

Given relations in the form presented in Sec.~\ref{sec:syzigies},  this can be done at lower degrees by the following simple approach (``by-hand''): one excludes those monomials that have as a factor any of the products appearing in LHS of the syzygies. We were able to follow this procedure and construct all the required bases except for that needed for $\T330$ (table \ref{tab:d_n_3}).   

Due to this, we switched to a more general technique\footnote{In principle, \Macaulay\ allows one to generate all basis elements of a certain degree in quotient rings with a single $\texttt{basis}$ command. However, we did not succeed in its application to $\mathcal{R}$ and reverted to the method described in the text.}. Given a Gr\"oebner basis $G$ for the ideal $\IdealTHDM$, we can test if a monomial $m_{\alpha\beta} \equiv g_1^{\alpha_1} \ldots g_M^{\alpha_M}$ having some degree $\{\alpha,\beta\}$  belongs to the corresponding basis of $\mathcal{R}_{\alpha\beta}$ by considering its \emph{normal} form $(m_{\alpha\beta}\mod G)$. If the element $m_{\alpha\beta}$ cannot be reduced in terms of other ring elements, its normal form coincides with the initial expression. In the opposite case, we exclude $m_{\alpha\beta}$ from the set 
$\{I^{(i)}_{\alpha,\beta}\}$.
In such a way we cross-checked the bases for $\beta<3$, and explicitly constructed all the linear-independent invariants for $\beta=3$. 
\subsection{\label{sec:RG_results_1l}Analytic expressions for beta functions}

Given all the needed elements 
$\{I^{(i)}_{\alpha,\beta}\}$, by straightforward but tedious algebra we derive the six-loop beta functions for all invariants from the generating set 
\begin{align}
	\beta_{g_n} \equiv \sum\limits_{l=1}^6 h^l \beta_{g_n}^{(l)}    
	\label{eq:beta_def_6L}
\end{align}
 To save space, we restrict ourselves to the one-loop results in the main text and refer to the supplementary material for the full expressions: 
\begin{align}
	\beta^{(1)}_{\Z11} & = 
	 28 \T200
	+48 \T002
	+\frac{29}{3} \Z11^2
	+2 \Z11 \Z12
	+3 \Z12^2,
	\label{eq:b1_z11}
	\\
\beta^{(1)}_{\Z12} & = 
	-12 \T200
	+\Z11^2
	+6 \Z11 \Z12
	+5 \Z12^2,
	\label{eq:b1_z12}
	\\
\beta^{(1)}_{\T200} & = 
	\frac{8}{3} \left[7 \Z11 - 3 \Z12\right] \T200 
	+ 96 \T300
	+24 \T102,
	\label{eq:b1_T200}
	\\
\beta^{(1)}_{\T300} & = 
	4 \left[7  \Z11 -3 \Z12\right] \T300
	+\frac{16}{3} \T200^2
	-4 \T202,
	\label{eq:b1_T300}
	\\
\beta^{(1)}_{\T002} & = 
	32 \Z11 \T002  
	+ 48 \T102
	\label{eq:b1_T002}
	\\
\beta^{(1)}_{\T102} & = 
	\frac{4}{3} \left[31  \Z11 -  3 \Z12\right] \T102  
	+ 16 [ 2\T200 + \T002] \T002
	-\frac{64}{3} \T202,
	\label{eq:b1_T102}
	\\
\beta^{(1)}_{\T202} & = 
	\frac{8}{3} \left[19 \Z11 -  3 \Z12\right] \T202
	-288 \T300 \T002
	-16 [5 \T200 + 3 \T002 ]\T102,
	\label{eq:b1_T202}\\
\beta^{(1)}_{\T303} & = 
	12 [6 \Z11 - \Z12] \T303
	\label{eq:b1_T303}
\end{align}
The one-loop beta functions for the invariants linear in mass:
\begin{align}
\beta^{(1)}_{\Y1} & = 
	2 [3 \Z11 + \Z12] \Y1
	+ 24 \T011,
	\label{eq:b1_y}
	\\
\beta^{(1)}_{\T011} & = 
	2 [9  \Z11 -  \Z12] \T011
	+ 36 \T111
	+6 \T002 \Y1,
	\label{eq:b1_T011}
	\\
\beta^{(1)}_{\T111} & = 
	\frac{2}{3}\left[41  \Z11 -9 \Z12\right]    \T111 
	-\frac{52}{3} \T211
	+8 [3 \T200 + 2 \T002]\T011  
	+6 \T102 \Y1,
	\label{eq:b1_T111}
	\\
\beta^{(1)}_{\T211} & = 
	\frac{10}{3} [11  \Z11 -3  \Z12 ]\T211 
	-4 [17 \T200 + 6 \T002] \T111,
	\nonumber
	\\
		    & -24 (9 \T300  +  \T102) \T011 +6 \T202 \Y1,
	\label{eq:b1_T211}
	\\
\beta^{(1)}_{\T112} & = 
\frac{2}{3} [65 \Z11 - 9 \Z12] \T112 -28 \T212,
\label{eq:b1_T112}
	\\
\beta^{(1)}_{\T212} & = 
	\frac{2}{3}[ 79 \Z11 - 15 \Z12] \T212 
	-\frac{4}{3} [11 \T200  -6 \T002 ] \T112
	+4 \T312,
	\label{eq:b1_T212}
	\\
\beta^{(1)}_{\T312} & = 
2 [ 31 \Z11 - 7 \Z12] \T312 
	+18 \T303 \Y1
	\nonumber\\
		    &
	+24 [ 3 \T300 + \T102]  \T112
	+24 [\T200 + 2 \T002] \T212.
	\label{eq:b1_T312}
\end{align}
The one-loop beta functions for invariants quadratic in mass:
\begin{align}
\beta^{(1)}_{\T020} & = 
	4 [\Z11 - \Z12] \T020
	+24 \T120
	+12 \T011 \Y1,
	\label{eq:b1_T020}
	\\
\beta^{(1)}_{\T120} & = 
	\frac{8}{3}[5 \Z11 -3 \Z12] \T120  
	-\frac{40}{3} \T220
	\nonumber\\
		    &
	+8[2 \T200 - \T002]\T020
	+12 \T111 \Y1
	+24 \T011^2,
	\label{eq:b1_T120}
	\\
\beta^{(1)}_{\T220} & = 
\frac{4}{3} [17 \Z11 - 9 \Z12] \T220 
        +12 \T211 \Y1
       -144 (\T011 \T111 +  \T300 \T020) \nonumber \\
		    &
        +48 (\T102 \T020  + \T002 \T120)
        -56 \T200 \T120,
	\label{eq:b1_T220}
	\\
\beta^{(1)}_{\T121} & = 
\frac{8}{3} [11 \Z11 - 3 \Z12] \T121 
	-6 \T112 \Y1
	-4 \T221,
	\label{eq:b1_T121}
	\\
\beta^{(1)}_{\T221} & = 
\frac{4}{3} [ 29 \Z11 - 9 \Z12] \T221 
	-6 \T212 \Y1
	-4 \T321,
	\nonumber\\
		    &
	-\frac{4}{3} [ 5 \T200 - 12 \T002]  \T121
	+24 \T011 \T112,
	\label{eq:b1_T221}
	\\
\beta^{(1)}_{\T321} & = 
	16 [3 \Z11 - \Z12] \T321
	-24 [ 3 \T300 + 2 \T102] \T121
	\nonumber\\
		    &
	-24 [\T200 + \T002] \T221
	+72 \T011 \T212
	+12 \T312 \Y1.
	\label{eq:b1_T321}
\end{align}
Finally, for the only CP-odd generator, cubic in the mass parameters, we have 
\begin{align}
\beta^{(1)}_{\T330} & = 
	2 (17 \Z11 - 9 \Z12) \T330 
	+24 [\T120 \T112 - \T020 \T212]
	-72 \T011 \T221
	+6 \T321 \Y1.
	\label{eq:b1_T330}
\end{align}

It is worth stressing that all the beta functions are written for a particular monomial ordering (see appendix~\ref{app:gb_syzygies}), i.e., in  the corresponding \emph{normal form}. This also means that the beta function $\beta_p$ for an arbitrary polynomial  $p(g_1,\ldots,g_M) \in \mathcal{R}$ is given by 
\begin{align}
	\beta_p \equiv \frac{d}{dt} p = \left[\sum\limits_{n=1}^{M} \frac{\partial p}{\partial g_n} \beta_{g_n}\right] \mod G.
	\label{eq:RG_for_poly}
\end{align}
We use this definition to explicitly prove that $K=63$ polynomials $\{r_k\}$, which generate $\IdealTHDM$, are RG-invariant up to six loops\footnote{See the supplementary material for a code in \Macaulay.}.

If one changes the monomial ordering, the Gr\"obner basis for the ideal $\IdealTHDM$ should be recalculated, and a new representative should be taken in the quotient ring $\mathcal{R}$. Nevertheless, this task can be easily automatized, e.g., by the \Macaulay\ package.

\begin{table}
	\centering
	\begin{tabular}{rcccccccccc}
      & $\T020$  & $\T120$ & $\T220$  &   &   &   &   &   &   &  \\
 \hline
 $\alpha$ & 1 & 2 & 3 & 4 & 5 & 6 & 7 & 8 & 9 & 10 \\
 \hline
 {even} & \bf{6} & \bf{17} & \bf{38} & \bf{78} & \bf{144} & \bf{254} & \bf{420} & \bf{671} & 1030 & 1539 \\
 {odd} & 0 & 1 & \bf{4} & \bf{12} & \bf{28} & \bf{60} & \bf{114} & \bf{205} & \bf{346} & \bf{561} \\
 \hline
      &    &  & $\T121$  & $\T221$  & $\T321$  &   &   &   &   &  
\end{tabular}
\caption{The number $d^{\text{even}}_{\alpha,2}$ ($d^{\text{odd}}_{\alpha,2}$) of CP-even (CP-odd) linear-independent invariants with multi degree $\{\alpha,2\}$. } 
\label{tab:d_n_2}
\end{table}

\begin{table}
	\centering
\begin{tabular}{rccccccccc}
 \hline
 ${\alpha}$ & 1 & 2 & 3 & 4 & 5 & 6 & 7 & 8 & 9 \\
 \hline
 {odd} & 0 & 1 & 7 & \bf{21} & \bf{53} & \bf{115} & \bf{225} & \bf{407} & \bf{697} \\
 \hline
  &   & &   & $\T330$   &   &   &   &   &   \\
\end{tabular}
\caption{The number $d^{\text{odd}}_{\alpha,3}$ of CP-odd linear-independent invariants with multi degree $\{\alpha,3\}$.} 
\label{tab:d_n_3}
\end{table}

\section{\label{sec:conclusios}Conclusions}

We have considered a polynomial ring $R$ of basis-independent quantities constructed from the parameters entering the scalar potential of the general 2HDM. The importance of these quantities comes from the fact that observables should not depend on the basis choice in the field configurations and, thus, are functions of the elements of the ring. 
In this paper we derived the six-loop RG equations for such elements in the limit of vanishing gauge and Yukawa interactions.  

Twenty two variables $\{g_n\}$ of $R$ constitute a generating set~\cite{Trautner:2018ipq}. 
The important subtlety is the existence of polynomial relations (syzygies) between the generators: we have only eleven algebraically independent invariants corresponding to the physical parameters of the Higgs sector. The syzygies form an ideal $\IdealTHDM$ in the ring, and one needs to consider all polynomials modulo these relations (conjugacy classes in the quotient ring $\mathcal{R}=R/\IdealTHDM$). 

The beta functions $\beta_{g_n}$ of $g_n$ are themselves elements of $\mathcal{R}$ and thus can be given in a \emph{normal form}, i.e., by a representative of their conjugacy class. We address this issue via the Gr\"obner-basis methods. By choosing various monomial orderings in $R$, we used the \Macaulay\ code to find a minimal set of relations that generate $\IdealTHDM$, and explicitly constructed relevant sets of linear-independent products of $g_n$ that can enter $\beta_{g_n}$ at each loop order. This information together with the known RG functions for the couplings and masses \cite{Bednyakov:2021ojn} allowed us to derive the expressions for $\beta_{g_n}$ by simple linear algebra.  

In due course of our calculation we heavily rely on the information encoded in the Hilbert series and the corresponding plethystic logarithms. As a by-product of our study we re-derived the explicit expression for the 2HDM Hilbert series by means of the so-called free resolution of the syzygy module. The latter, in our opinion, provides more insights into the first positive and negative terms that appear in the expansion of plethystic logarithms.

We provide our results in a computer-readable form together with the simple scripts in \Mathematica\ and \Macaulay\ allowing one to compute the scale dependence of any polynomial from $\mathcal{R}$, or use the obtained expressions, e.g., in a study of various symmetries of 2HDM \cite{Bento:2020jei,Ferreira:2023dke} in a basis-independent way.  

A possible continuation of the study is to take into account gauge and Yukawa interactions relevant, e.g., to the recently discovered non-conservation of CP in the real 2HDM \cite{deLima:2024hnk,deLima:2024lfc}. One can also extend the approach to multi-doublet Higgs models (NHDM) \cite{Bento:2021hyo}.

\section{Acknowledgements}
	The author would like to thank  Alexey Isaev, Roman Lee, Andrey Onishchenko, and Nikolai Tyurin for fruitful discussions. 
	We are also grateful for the numerous comments and suggestions from anonymous reviewers.
	\appendix
	\section{\label{app:monomial_ordering_and_gb}Monomial ordering, polynomial division, ideals and Gr\"oebner bases}
	Here we give a brief introduction to Gro\"ebner bases and their primary application related to ideal membership. More detail can be found in ref.~\cite{CoxLittleOShea}. 

	Let us consider a polynomial ring $R=k[y_1,\ldots,y_n]$ in $n$ variables over some field (or ring) $k$. The elements of $R$ are linear combinations of monomials $y^\alpha = y_1^{\alpha_1}\ldots y_n^{\alpha_n}$ with exponents defined by an $n$-tuple $(\alpha_1,\ldots,\alpha_n)\in \mathbb{Z}^n_{\geq0}$. A useful notion is the monomial ordering allowing one to say that the monomial $y^\alpha$ is ``higher'' than $y^\beta$, i.e., $y^\alpha > y^\beta$. The one-to-one correspondence between a monomial $y^\alpha$ and the exponent $\alpha$ allows one to define 
	an ordering by comparing exponents. There are many ways to do this \cite{CoxLittleOShea}.

	For example, in \Macaulay, which we use in our study, the default ordering is graded reverse lexicographical (GRevLex):  one says that $\alpha > \beta$ if either $|\alpha|=\sum_i\alpha_i > |\beta|=\sum_i\beta_i$, or $|\alpha|=|\beta|$ and the rightmost non-zero entry of $\alpha - \beta \in \mathbb{Z}^n$ is negative. According to this ordering, e.g.,  $y_1^3 y_2^3 > y_1^3 y_2^2$, since $|\alpha|=6>|\beta|=5$ but $y_1^3 y_2^2 > y_1^2 y_2^3$, since $|\alpha|=|\beta|=5$ and we have a negative rightmost element in $\alpha - \beta = (1,-1)$ that is nonzero.  

	One can also define a (partial) ordering by introducing an integer \emph{weight} vector $\omega = (\omega_1\ldots\omega_n)$: 
	$y^\alpha > y^\beta$ if $\alpha \cdot \omega > \beta \cdot \omega$. The \Macaulay\ package allows us to specify $\omega$ for the ring variables and revert to GRevLex in case $\alpha \cdot \omega = \beta \cdot \omega$. We use a deliberately chosen weight $\omega$ to obtain some of the results presented in appendix~\ref{app:gb_syzygies}. 

Given the monomial ordering, one uniquely defines the leading term $\LT(p)$ of a polynomial $p$ with respect to this ordering. 
 The leading terms play an important role in the polynomial division algorithm \cite{CoxLittleOShea}. 

 When we divide a polynomial $p_1$ over a polynomial $d$
, we find a quotient $q_1$ and a remainder $r_1$ in the representation $p_1 = q_1 \cdot d + r_1$. 
The leading term $\LT(q_1)$ is determined from the requirement $\LT(p_1) = \LT(q_1) \LT(d)$. As a consequence,  $\LT(p_1)$ is canceled in $p_2 \equiv [p_1 - \LT(q_1) d]$. 
The next term in $q_1$, or equivalently $\LT(q_2)$ for $q_2 \equiv q_1  - \LT(q_1)$, is found by the same procedure, e.g., $\LT(p_2) = \LT(q_2) \LT(d)$, and $p_3 \equiv [p_2 - \LT(q_2) d]$ is free from $\LT(p_1)$. Repeating the latter, we will find at a certain step that either $p_i=0$ corresponding to $r_1 = 0$, or $\LT(p_i)$ is not divisible by $\LT(d)$. 
In the latter, case we determine the leading term of the remainder, $\LT(r_1) = \LT(p_i)$, introduce $r_2 \equiv r_1 - \LT(p_i)$, 
and continue with $p_{i+1} \equiv p_i - \LT(p_i)$ until at some step we eventually get $p_{i+m} = 0$ and stop.
In such a way we obtain all the terms of $q_1 = \LT(q_1) + \LT(q_2) + \ldots$ and $r_1 = \LT(r_1) + \LT(r_2) + \ldots$. One sees that none of the terms in $r_1$ is divisible by 
$\LT(d)$.

The division algorithm can be extended to the case of several divisors $d_i$, i.e., a polynomial $p$ belonging to some ring $R$ can be represented as
\begin{align}
    p = q_1 d_1 + \ldots + q_n d_n + r
    \label{eq:poly_division}
\end{align}
with the quotients $q_i \in R$ and the remainder $r \in R$. 

The representation \eqref{eq:poly_division} is crucial when considering ideals in polynomial rings, or quotient rings. 
If we have an ideal $\mathcal{I} = \langle d_1\ldots d_n\rangle$, and for a polynomial $p$ in \eqref{eq:poly_division} we obtain $r=0$, then $p\in \mathcal{I}$. The condition $r=0$ is sufficient to determine whether $p$ belongs to the ideal, but not necessary: for a fixed set $\{d_i\}$ both $q_i$ and $r$ are not unique 
and depend on the order of division.\footnote{For example \cite{CoxLittleOShea},
$x^2 y + x y^2 + y^2$ can be represented as $(x+y) f_1 + f_2 + [x + y + 1]$ or $ (x+1) f_2 + x f_1 + [2 x + 1]$ for $f_1 = x y -1$ and $f_2 = y^2 -1$, depending on whether we divide first by $f_1$ or $f_2$.}  This issue can be circumvented by construction of the \emph{reduced} Gr\"obner basis $G = \langle \mathfrak{g}_1\ldots \mathfrak{g}_m\rangle$ for $\mathcal{I}$, i.e a different \emph{uniquely} determined generating set $\{\mathfrak{g}_n\}$, such that the remainder $r$ (or the \emph{normal form} $(p\mod G)$ of $p$ in the quotient ring $R/\mathcal{I}$) is unique and the condition $r=0$ is equivalent to the membership of the ideal. 

We will not discuss here the algorithms (i.e., due to Buchberger) allowing one to construct a Gr\"ober basis but refer to the literature \cite{CoxLittleOShea,DerksenKemper}. In the current study we utilize special software to carry out this tedious task.

\section{\label{app:gb_syzygies}Algorithm to compute syzygies}

Let us briefly describe the algorithm to find the generators of $\IdealTHDM$ and discuss its implementation in  \Macaulay. 
The idea (see, e.g., ref.~\cite{CoxLittleOShea}) is to consider a graded polynomial ring\footnote{In this appendix we consistently exclude 3 singlets of degree 1 from $\{g_n\}$, thus, assuming $M=19$ here.}  $\tilde R = \mathtt{k}[x_1,\ldots,x_N,g_1,\ldots,g_M]$. 
The first $N$ variables $x_1\ldots x_N$ have degree 1 and are linear in couplings and masses. They parametrize building blocks used to construct the basis invariants\footnote{Note that \eqref{eq:g_parametrization} defines the map $\phi$ discussed in Sec.~\ref{sec:polynomial_ring}.} $g_n$ defined in Sec.~\ref{sec:generating_set}
\begin{align}
	g_n = g_n(x_1,\ldots,x_N), \qquad n=1\ldots M.
	\label{eq:g_parametrization}
\end{align}
The degree of $g_n$ as a variable can be easily computed from RHS \eqref{eq:g_parametrization}, since the latter are homogeneous functions of $x_i$. For example, the degree of $\T200$ is 2, while $\T303$ has degree 6.  

In our study we used two parametrizations\footnote{See the supplementary material.} to derive the syzygies. The first one is in terms of $N=11$ variables 
\begin{align}
	\{\underbrace{y, y_r, y_i}_{\text{mass triplet}}, 
	  \underbrace{t, t_r, t_i}_{\text{coupling triplet}}, 
  \underbrace{q_{r1}, q_{i1}, q_{r2},q_{i2}, q_3}_{\text{coupling fiveplet}}\},
\label{eq:Trautner_pars}
\end{align}
given in appendix D.1 of ref.~\cite{Trautner:2018ipq}. 

The second parametrization is based on the expressions given in Section~\ref{sec:generating_set}. 
Under the assumption that we have a non-degenerate $\lamT$, one can diagonalize the latter. In this basis, we introduce $N=8$ free parameters\footnote{Adding 3 degree-one singlets we get the total number of physical parameters of 2HDM.}  
\begin{align}
	\tilde\Lambda = \text{diag}(l_1, l_2, -l_1- l_2), \quad \lamV = (v_1, v_2, v_3), \quad \mV = (m_1, m_2, m_3).
	\label{eq:so3_diagonal_lambda_pars}
\end{align}
The algorithm is independent of parametrization. Both these choices give rise to the same syzygies. However, the \Macaulay\ computer code utilizing \eqref{eq:so3_diagonal_lambda_pars} is about 20 times faster than that\footnote{Both \Macaulay\ scripts are available as supplementary files}  based on \eqref{eq:Trautner_pars}. 

To find syzygies, we consider the ideal generated by the polynomial $g_n - g_n(x_1,\ldots,x_N)$ in the ring $\tilde R$
\begin{align}
	\mathcal{I} = \langle g_1 - g_1(x_1,\ldots,x_N), \ldots, g_M - g_M(x_1,\ldots,x_N)\rangle
\end{align}
and define $\IdealTHDM$ as an \emph{elimination} ideal 
\begin{align}
	\IdealTHDM = \mathcal{I} \cap \mathbb{R}[g_1,\ldots,g_M] 
\end{align}
which consists of all consequences of $g_n - g_n(x_1,\ldots,x_N)=0$ that are free from $x_i$.  The subtlety here is that we have an infinite number of such relations in $\IdealTHDM$. 

To deal with this problem, we define partial ordering in $\tilde R$ by introducing the weight vector 
\begin{align}
\omega_1 = (\underbrace{1\ldots 1}_N,\underbrace{0\ldots0}_M) 
\label{eq:omega_in_R_tilde}
\end{align}
that makes any monomial involving at least one variable $x_i$ ``higher'' than that depending solely on $g_n$. 
For example, given parameterization in terms of \eqref{eq:so3_diagonal_lambda_pars}, the graded ring $\tilde R$ and the ideal $\mathcal{I}$ are defined in \Macaulay\ via the code

\begin{lstlisting}[language={},caption={}]
-- define a ring over rationals QQ

R = QQ[l1,l2,l3,m1,m2,m3, T200, T300, ... , T330, T303,
      Degrees => 
      {1, 1 ,1, 1 ,1 ,1 , 2   , 3   , ... , 6   , 6   },
      MonomialOrder => Eliminate 8];

-- define an ideal

I = ideal( 
        T200 - 1/4*(l1^2 + l2^2 + l1*l2),
        T300 + 1/16*(l1^2*l2 + l1*l2^2),
        ...
        T330 - ... ,
        T303 - ...
	)

\end{lstlisting}
where for brevity we omit the unnecessary detail. The \texttt{Degree} option specifies the (total) degrees of the variables, 
while \texttt{MonomialOrder} allows one to specify the ordering. The keyword \texttt{Eliminate 8} precisely corresponds to the weight vector $\omega_1$ with $N=8$. 
Let us mention here that in order to speed up the computation, we use rationals $\mathbb{Q}$ thus omitting the imaginary unit in the definition of CP-odd invariants. 
We restore the correct signs at the final stage (see below).  

To find the generators of $\IdealTHDM$, we construct a (partial) Gr\"obner basis for $\mathcal{I}$ by restricting ourselves to the total degree 12. 
\Macaulay\ allows one to automatically extract the syzygies belonging to $\IdealTHDM$ via $\texttt{selectInSubring}$ command applied to the generators (\texttt{gens}) of the Gr\"obner basis: 

\begin{lstlisting}[language={},caption={}]
-- constuct Groebner basis 
GB = gb(I,DegreeLimit => 12)

-- extact relations between Tabc only
GBrel = selectInSubring(1,gens GB)

\end{lstlisting}
The restriction \texttt{DegreeLimit => 12} comes from the highest degree of the ``first negative terms'' in plethystic logarithm \eqref{eq:PL_unigraded}. 
It turns out that the computation of higher-degree relations is very time-consuming, yet no new information is generated.
Moreover, the relations stored in \texttt{GBrel} are not minimal generators of $\IdealTHDM$. 

To find $K=63$ minimal generators, we define a new (multigraded) ring $\mathtt{k}[g_1, \ldots, g_M]$ over a ``field of complex numbers'' $\mathtt{k}$ :
\begin{lstlisting}[language={},caption={}]
-- field of complex numbers represented as a quotient ring 
k = toField QQ[i]/(i^2+1)

RI = k[  T200,    T300, ...,   T112,  , ...,   T330,    T303,
      Degrees => {
      {2,0,0}, {3,0,0}, ..., {1,1,2}  , ..., {3,3,0}, {3,0,3}},
      Weights => {
      {     0,       0, ...,       5  , ...,       7,       6}
      ];

\end{lstlisting}
Here we use an auxiliary weight vector $\omega_2$ that assigns the weights to the variables given in table.~\ref{tab:weights}.
\begin{table}
	\centering
	\begin{tabular}{c|c}
		weight & invariants \\
		\hline
		0 & \T200, \T300, \T020, \T011, \T002, \T120, \T102, \T202\\ 
		1 & \T220 \\
		2 & \T211 \\
		3 & \T111 \\
		5 & \T112, \T121, \T212, \T221, \T312, \T321\\
		6 & \T303\\
		7 & \T330
	\end{tabular}
	\caption{A variant ofthe  weight vector $\omega_2$ that gives rise to $K=63$ syzygies 
	with the leading terms given in Sec.~\ref{sec:syzigies}. 
	The terms involving higher weights are to be eliminated first. } 
	\label{tab:weights}
\end{table}

To define the ideal $\IdealTHDM$, we restore the imaginary units in the definition of CP-odd invariants and compute its minimal generators $\{r_k\}$ via:
\begin{lstlisting}[language={},caption={}]
-- replace the initial ring R by RI in the relations GBrel
GBrel = sub(GBrel, RI);

-- restore imaginary units in CP-odd invariants 
GBrel = sub(GBrel, {
                    T112 => -i*T112,
                    ...
                    T330 => -i*T330,
                    T303 => -i*T303
                   });
		
-- define the Ideal2HDM by relations from GBrel
II = ideal(GBrel)

-- compute syzygies as minimal generators of II
syzygies = mingens (II) 
\end{lstlisting}
The obtained syzygies can be used subsequently to define the quotient ring $\mathcal{R} = R/\IdealTHDM$ of basis invariants.

\section{\label{app:Hilbert_series}Free resolution of the syzygy module and Hilbert Series} 

The fact that any element of $\IdealTHDM$ can be represented as a linear combination of its generators allows us to treat it as a \emph{finitely generated $R$-module} $\ModuleTHDM$ with generators $\{r_k\}$. Similar to vector spaces, given the generators, we can specify any element $f \in \ModuleTHDM$ by its coordinates $\{a_k\}$, i.e., by an element belonging to Cartesian product $R^K = \underbrace{R \times R \times \ldots \times R}_{K \text{ times}}$: 
\begin{align}
	f = \sum\limits_{k=1}^K a_k \cdot r_k  \qquad a_k \in R.
	\label{eq:syz_module}
\end{align}
Let us consider a map $\phi_0: R^K \to R$ that maps an $M$-tuple $\{a_k\}$ to the corresponding element $f\in R$  belonging to $\IdealTHDM$ via \eqref{eq:syz_module}, i.e., $\IdealTHDM = \text{im } \phi_0$.  
It turns out that there are relations (higher syzygies) of the form\footnote{An obvious example is $(r_i) r_j - (r_j) r_i = 0$ for fixed $i,j$, where the first factors are treated as ones belonging to ${R}$, while the second ones as generators of $\ModuleTHDM$.}
\begin{align}
	\sum\limits_{k=1}^K b_k \cdot r_k = 0 
	, \qquad  b_k \in R.
	\label{eq:higher_syz}
\end{align}
Speaking differently, $\ker \phi_0$ is non-trivial and $\ModuleTHDM$ is not \emph{free}, i.e, not isomorphic to a free $R$-module $F_0 \equiv R^K$. 
The set of  $K$-tuples $\{b_k\}$ satisfying \eqref{eq:higher_syz} can again be treated as an $R$-module $\mathcal{M}_1 = \ker \phi_0$: we can add such tuples and multiply by elements of $R$.
Due to this, we have $\ModuleTHDM=F_0/\mathcal{M}_1$. 

As in the case of $\ModuleTHDM$, one introduces a generating set $b^{(s)} = \{b_1^s,\ldots b_K^s\}$ of $\mathcal{M}_1$ with $s=1\ldots S_1$. Again any element $f_1 \in \mathcal{M}_1$ can be represented by an $S_1$-tuple $\{c_s\}$ 
\begin{align}
	f_1 = \sum\limits_{s=1}^{S_1} c_s \cdot b^{(s)}, \qquad  c_s \in R
\end{align}
and one defines a \emph{linear} map $\phi_1: F_1 \equiv R^{S_1} \to F_0$ that takes an $S_1$-tuple $\{c_s\}$ and returns a $K$-tuple $\{a_k\}$ via 
\begin{align}
	a_k = b^{(s)}_k c_s , \qquad \{a_k\} \in R^k, \quad \{c_s\} \in R^{S_1}. 
	\label{eq:linear_map_modules}
\end{align}
As a consequence, one sees that $\mathcal{M}_1 = \text{im} \, \phi_1$. 
The situation is similar to that used to define $\IdealTHDM$. The module $\mathcal{M}_1$ turns out to be again non-free, so that $\mathcal{M}_1=F_1/\mathcal{M}_2$ with $\mathcal{M}_2 = \ker \phi_1$.
According to Hilbert's syzygy theorem (see, e.g., ref.~\cite{DerksenKemper}), repeating this procedure  we eventually obtain a free module $F_J$ at step $J$ such that $\ker \phi_J = 0$. This completes  the free resolution of the finitely generated module $\ModuleTHDM$. 
The process can be conveniently represented by an exact sequence ($\ker \phi_i = \text{im}\, \phi_{i+1}$):
\begin{align}
	0 \stackrel{\phi}{\longleftarrow} R \stackrel{\phi_0}{\longleftarrow} F_0 \stackrel{\phi_1}{\longleftarrow} F_1 \stackrel{\phi_2}{\longleftarrow} F_2 \longleftarrow \ldots
	\stackrel{\phi_J}{\longleftarrow}  F_J \longleftarrow 0. 
    \label{eq:free_res_ex}
\end{align}
By means of \Macaulay\ we can construct such a resolution explicitly\footnote{With sufficient amount of CPU power, computer memory, and time.}, i.e., find generators for all the modules and maps between the latter. 

\begin{table}[t]
	\begin{center}
{\tiny
\begin{tabular}{c|rrrrrrrrrrrr}
  & 0 & 1 & 2 & 3 & 4 & 5 & 6 & 7 & 8 & 9 & 10 & 11 \\
\hline
  \text{total} & 1 & 63 & 411 & 1358 & 2835 & 4038 & 4038 & 2835 & 1358 & 4111 & 63 & 1 \\
\hline
 0 & 1 & 0 & 0 & 0 & 0 & 0 & 0 & 0 & 0 & 0 & 0 & 0 \\
\ldots   & \multicolumn{11}{c}{\ldots} \\
 5 & 0 & 1 & 0 & 0 & 0 & 0 & 0 & 0 & 0 & 0 & 0 & 0 \\
 6 & 0 & 3 & 0 & 0 & 0 & 0 & 0 & 0 & 0 & 0 & 0 & 0 \\
 7 & 0 & 12 & 0 & 0 & 0 & 0 & 0 & 0 & 0 & 0 & 0 & 0 \\
 8 & 0 & 12 & 2 & 0 & 0 & 0 & 0 & 0 & 0 & 0 & 0 & 0 \\
 9 & 0 & 17 & 14 & 0 & 0 & 0 & 0 & 0 & 0 & 0 & 0 & 0 \\
 10 & 0 & 8 & 38 & 0 & 0 & 0 & 0 & 0 & 0 & 0 & 0 & 0 \\
 11 & 0 & 10 & 63 & 1 & 0 & 0 & 0 & 0 & 0 & 0 & 0 & 0 \\
 12 & 0 & 0 & 91 & 23 & 0 & 0 & 0 & 0 & 0 & 0 & 0 & 0 \\
 13 & 0 & 0 & 79 & 60 & 0 & 0 & 0 & 0 & 0 & 0 & 0 & 0 \\
 14 & 0 & 0 & 72 & 143 & 0 & 0 & 0 & 0 & 0 & 0 & 0 & 0 \\
 15 & 0 & 0 & 32 & 220 & 16 & 0 & 0 & 0 & 0 & 0 & 0 & 0 \\
 16 & 0 & 0 & 20 & 274 & 56 & 0 & 0 & 0 & 0 & 0 & 0 & 0 \\
 17 & 0 & 0 & 0 & 260 & 170 & 0 & 0 & 0 & 0 & 0 & 0 & 0 \\
 18 & 0 & 0 & 0 & 196 & 307 & 4 & 0 & 0 & 0 & 0 & 0 & 0 \\
 19 & 0 & 0 & 0 & 118 & 484 & 31 & 0 & 0 & 0 & 0 & 0 & 0 \\
 20 & 0 & 0 & 0 & 48 & 543 & 107 & 0 & 0 & 0 & 0 & 0 & 0 \\
 21 & 0 & 0 & 0 & 15 & 527 & 265 & 0 & 0 & 0 & 0 & 0 & 0 \\
 22 & 0 & 0 & 0 & 0 & 370 & 484 & 8 & 0 & 0 & 0 & 0 & 0 \\
 23 & 0 & 0 & 0 & 0 & 234 & 682 & 33 & 0 & 0 & 0 & 0 & 0 \\
 24 & 0 & 0 & 0 & 0 & 92 & 788 & 136 & 0 & 0 & 0 & 0 & 0 \\
 25 & 0 & 0 & 0 & 0 & 32 & 697 & 280 & 0 & 0 & 0 & 0 & 0 \\
 26 & 0 & 0 & 0 & 0 & 4 & 523 & 523 & 4 & 0 & 0 & 0 & 0 \\
 27 & 0 & 0 & 0 & 0 & 0 & 280 & 697 & 32 & 0 & 0 & 0 & 0 \\
 28 & 0 & 0 & 0 & 0 & 0 & 136 & 788 & 92 & 0 & 0 & 0 & 0 \\
 29 & 0 & 0 & 0 & 0 & 0 & 33 & 682 & 234 & 0 & 0 & 0 & 0 \\
 30 & 0 & 0 & 0 & 0 & 0 & 8 & 484 & 370 & 0 & 0 & 0 & 0 \\
 31 & 0 & 0 & 0 & 0 & 0 & 0 & 265 & 527 & 15 & 0 & 0 & 0 \\
 32 & 0 & 0 & 0 & 0 & 0 & 0 & 107 & 543 & 48 & 0 & 0 & 0 \\
 33 & 0 & 0 & 0 & 0 & 0 & 0 & 31 & 484 & 118 & 0 & 0 & 0 \\
 34 & 0 & 0 & 0 & 0 & 0 & 0 & 4 & 307 & 196 & 0 & 0 & 0 \\
 35 & 0 & 0 & 0 & 0 & 0 & 0 & 0 & 170 & 260 & 0 & 0 & 0 \\
 36 & 0 & 0 & 0 & 0 & 0 & 0 & 0 & 56 & 274 & 20 & 0 & 0 \\
 37 & 0 & 0 & 0 & 0 & 0 & 0 & 0 & 16 & 220 & 32 & 0 & 0 \\
 38 & 0 & 0 & 0 & 0 & 0 & 0 & 0 & 0 & 143 & 72 & 0 & 0 \\
 39 & 0 & 0 & 0 & 0 & 0 & 0 & 0 & 0 & 60 & 79 & 0 & 0 \\
 40 & 0 & 0 & 0 & 0 & 0 & 0 & 0 & 0 & 23 & 91 & 0 & 0 \\
 41 & 0 & 0 & 0 & 0 & 0 & 0 & 0 & 0 & 1 & 63 & 10 & 0 \\
 42 & 0 & 0 & 0 & 0 & 0 & 0 & 0 & 0 & 0 & 38 & 8 & 0 \\
 43 & 0 & 0 & 0 & 0 & 0 & 0 & 0 & 0 & 0 & 14 & 17 & 0 \\
 44 & 0 & 0 & 0 & 0 & 0 & 0 & 0 & 0 & 0 & 2 & 12 & 0 \\
 45 & 0 & 0 & 0 & 0 & 0 & 0 & 0 & 0 & 0 & 0 & 12 & 0 \\
 46 & 0 & 0 & 0 & 0 & 0 & 0 & 0 & 0 & 0 & 0 & 3 & 0 \\
 47 & 0 & 0 & 0 & 0 & 0 & 0 & 0 & 0 & 0 & 0 & 1 & 0 \\
\ldots   & \multicolumn{11}{c}{\ldots} \\
 52 & 0 & 0 & 0 & 0 & 0 & 0 & 0 & 0 & 0 & 0 & 0 & 1 
\end{tabular}	
}
\end{center}
\caption{Betti diagram obtained with \Macaulay{} \cite{M2} corresponding to the free resolution of $\ModuleTHDM$ \eqref{eq:free_resolution}.
The columns correspond to homological degree, while row labels give (multi) degree.
Each element $b_{ij}$ corresponding to row label $i$ and column label $j$ 
gives rise to a term $(-)^j b_{ij} t^{i+j}$ in the numerator $\tilde N(t)$ defined in Eq.~\eqref{eq:HS_from_resolution}.}
\label{tab:betti_table}
\end{table}

An application of such a resolution is the computation of the Hilbert series from the corresponding \emph{Betti} diagram (given in table~\ref{tab:betti_table}). The Betti table is related to the free resolution of $\ModuleTHDM$: 
{\footnotesize
\begin{align}
	R 
\leftarrow R^{63} 
\leftarrow R^{411} 
\leftarrow R^{1358} 
\leftarrow R^{2835} 
\leftarrow R^{4038} 
\leftarrow R^{4038} 
\leftarrow R^{2835} 
\leftarrow R^{1358} 
\leftarrow R^{411} 
\leftarrow R^{63} 
\leftarrow R 
\leftarrow 0,
\label{eq:free_resolution}
\end{align}
}
in which we omit the first map $\phi$. Given table~\ref{tab:betti_table}, we can construct the numerator $\tilde N(t)$ of the Hilbert series \eqref{eq:HS_ungraded} defined as
\begin{align}
	\tilde N(t) = H(t) 
	&\times \left(1-t^1\right)^3 &  [Y_1, Z_{1_{(1)}}, Z_{1_{(2)}}] \\
	&\times \left(1-t^2\right)^4 & [ \T200,\T020,\T002,\T011]  \\
	&\times \left(1-t^3\right)^4 & [ \T300,\T120,\T102,\T111]\\
	& \times \left(1-t^4\right)^5 & [ \T220,\T202,\T211,\T112, \T121]\\
	& \times \left(1-t^5\right)^2 & [ \T212,\T221]\\
	& \times \left(1-t^6\right)^4. & [ \T312,\T321,\T303,\T330]
\label{eq:HS_from_resolution}
\end{align}
Clearly, $\tilde N(t)$ is a polynomial in $t$. If we found $\tilde N(t)=1$, it would mean that there are no relations between the elements of the generating set. In this case, the number of invariants at a certain degree can be easily obtained by expanding factors $1/(1-t^{d_n})$ corresponding to the elements $g_n$ of the generating set with degree $d_n$. Non-trivial $\tilde N(t)\neq 1$ indicates that there are relations in the ring and, possibly, relations between relations.

Each element $b_{ij}$ corresponding to the row label $i$ and column label $j$ 
gives rise to a term $(-)^j b_{ij} t^{i+j}$ in the numerator $\tilde N(t)$~\eqref{eq:HS_from_resolution}. 
For example, the second column enumerates $K=63$ first syzygies and correctly reproduces their degrees: there are $b_{51}=1$ degree-6 relation, $b_{6,1}=3$ degree-7 relations, etc. All the elements with the column label $j=1$ contribute to $\tilde N(t)$ with a negative sign, reflecting the fact that we will overcount the invariants at the corresponding degrees if we neglect the relations (first syzygies) between the elements of the generating set. 
However, when going to higher degrees, there can be relations between first syzygies, so we would ``subtract'' too many invariants. This is fixed step-by-step by the add-and-subtract procedure in accord with the free resolution of $\ModuleTHDM$. For example,  there are $S_1=411$ linear relations \eqref{eq:higher_syz} between first syzygies enumerated in column $j=2$. The element $b_{8,2}=2$ gives rise to a term $+2 t^{10}$ corresponding to two degree-10 relations between $r_k$. This term compensates oversubtraction at degree 10 leading to $-[17 - 2]t^{10}$ in $\tilde N(t)$. 
Here we do not present the explicit expression for the numerator 
\begin{align}
	\tilde N(t) & = 1 - t^6 - 3 t^7 - 12 t^8 - 12 t^9 
	+ \ldots  
	+ 12 t^{54} 
	+ 12 t^{55} 
	+ 3 t^{56}
	+   t^{57}
	-   t^{63}, 
	\label{eq:numerator_free_res_exp}
\end{align}
satisfying the property 
\begin{align}
\tilde N(t) = - t^{63} \tilde N(1/t), 
\end{align}
but refer again to the supplementary material. 
One sees that the terms in $\tilde N(t)$ look very similar to those in plethystic logarithm \eqref{eq:PL_unigraded}. This fact can be explained by the following considerations.

Let us assume that the Hilbert series can be cast into a finite product (the standard Euler product form coined in ref.~\cite{Benvenuti:2006qr}) 
\begin{align}
	H(t) = \prod\limits_{n=1}^{\mathcal{N}} \frac{1}{(1-t^{a_n})^{d_n}},
	\label{eq:HS_normal_form}
\end{align}
for some finite $\mathcal{N}$ with natural $a_n>0$ and integer $d_n$. The latter can be both positive \emph{and negative}. The corresponding plethystic logarithm turns out to be a finite polynomial 
\begin{align}
	PL[H(t)] & =  \sum\limits_{k=1}^\infty  \frac{\mu(k)}{k} 
	\ln \left[\prod\limits_{n=1}^{\mathcal{N}} \frac{1}{(1-t^{k a_n})^{d_n}}\right] \\
		 & = 
		 -\sum\limits_{k=1}^\infty \sum\limits_{n=1}^\mathcal{N} \frac{\mu(k)}{k} 
	d_n \ln \left[1-t^{k a_n}\right]  = 
\sum\limits_{k=1}^\infty \frac{\mu(k)}{k} 
\left[-\sum\limits_{n=1}^\mathcal{N} 
d_n \ln \left[1-t^{k a_n}\right] \right]\\
		 & =   
		 \sum\limits_{n=1}^\mathcal{N} d_n \left[ 
\sum\limits_{k,m=1}^\infty \frac{\mu(k) (t^{a_n})^{k m} }{k m}\right]
= 
\sum\limits_{n=1}^{\mathcal{N}} d_n  
\sum\limits_{m=1}^\infty \left[\sum\limits_{k|m} \mu(k)\right] \frac{(t^{a_n})^{m} }{m} \\
		 & = 
		 \sum\limits_{n=1}^\mathcal{N} d_n  
\sum\limits_{m=1}^\infty \delta_{m,1} \frac{(t^{a_n})^{m} }{m} = 
\sum\limits_{n=1}^{\mathcal{N}} d_n t^{a_n},
\label{eq:PL_normal_form}
\end{align}
where the sum $\sum_{k|m}$  runs  over the divisors $k$ of $m$, and we use the properties of the M\"obius function $\sum_{k|m} \mu(k) = \delta_{m,1}$ (see, e.g., ref.~\cite{Feng:2007ur}). As a consequence, there is a one-to-one correspondence between the terms $d_n t^{a_n}$ in Eq.~\eqref{eq:PL_normal_form} and the factors in Hilbert series given in the form \eqref{eq:HS_normal_form}.  

It turns out that the plethystic logarithm for the full 2HDM invariant ring \eqref{eq:PL_unigraded} is an infinite series\footnote{It is said that it is not a complete intersection.}. Nevertheless, if we restrict ourselves to a certain finite degree, it can be approximated by Eq.~\eqref{eq:PL_normal_form} with some finite $\mathcal{N}$. 
The first positive terms in \eqref{eq:PL_unigraded} give rise to the same factors in $H(t)$ that we factor out to define $\tilde N(t)$ in \eqref{eq:HS_from_resolution}. Given other terms in \eqref{eq:PL_unigraded}, we can approximate
\begin{align}
	\tilde N(t)  \simeq &\frac{ 
	(1-t^6) 
        (1-t^7)^3 
        (1-t^8)^{12} 
        (1-t^9)^{12} 
	(1-t^{10})^{17}
	(1-t^{11})^{8}
	(1-t^{12})^{10}
}{
	(1-t^{10})^2
	(1-t^{11})^{14}
	(1-t^{12})^{38}
}
\\
=  & 1 - t^6 - 3 t^7 - 12 t^8 - 12 t^9 - [17 -2 ] t^{10} - [8-14] t^{11} - [10 - 38] t^{12} + \mathcal{O}(t^{13}) \nonumber
\end{align}
and, thus, correctly reproduce the exact expression for $\tilde N(t)$ up to degree $t^{13}$.

We close this appendix by mentioning that a multidegree version of the Betti table can in principle be constructed.

\section{\label{app:algebraic_dependence}Generating set and algebraically independent invariants}

By means of $\Macaulay$ \cite{M2} we were able to derive the algebraic relations of the form \eqref{eq:algebric_relation} for all the elements $g_n$ 
not belonging to the set of algebraically independent invariants \eqref{eq:alg_indep_invs_unigraded}. The key idea is to consider the syzygies $\{r_k\}$ and  
find an elimination ideal $\mathcal{I}_g = \IdealTHDM \cap \mathbb{R}[g,f_1,\ldots,f_N]$
for an invariant $g\in \{g_m\}$ and $g \notin \{f_i\}$. 
It turns out that for any 
\begin{align}
	\mathcal{J} \in \{\T111,\T202,\T220,\T112^2, \T121^2,\T212^2,\T221^2,\T312^2,\T321^2,\T303^2,\T330^2\}
\end{align}
we have a degree-four polynomial relation in $\mathcal{J}$: 
\begin{align}
	\sum\limits_{i=0}^4 \alpha^{(\mathcal{J})}_{i} \mathcal{J}^i  = 0,
	\qquad \alpha^{(\mathcal{J})}_i \in \mathbb{Z}[\T200,\T002,\T020,\T011,\T300,\T102,\T120,\T211],
\end{align}
where the coefficients $\alpha^{\mathcal{J}}_i$ are polynomials in invariants from the algebraically independent set. We provide the full form of the relations as supplementary material. Here we presents them schematically and indicate the corresponding multidegree: 
\begin{align}
	\T111: && [9 \times \,\T102 \T120]\cdot [\,\T111\,\,]^4 & + \ldots  = 0, & [6,6,6] 
	\label{eq:T111_ideal}\\
	\T202: && [3 \times \T120^2]\cdot [\,\T202\,\,]^4 & + \ldots = 0, & [10,4,8] 
	\label{eq:T202_ideal}\\
	\T112: && [729 \times \T102^2 \T120^2]\cdot [\T112^2]^4 & + \ldots = 0, & [12 , 12 , 20 ] 
	\label{eq:T112_ideal}\\
	\T212: && [6561\times \T120^4]\cdot [\T212^2]^4& + \ldots = 0, & [20,16,16 ] 
	\label{eq:T212_ideal}\\
	\T312: && [729 \times \T102^2 \T120^2] \cdot [\T312^2]^4 & + \ldots = 0, & [28 ,12 ,20 ] 
	\label{eq:T312_ideal}\\
	\T303: && [531441 \times \T102^6] \cdot[\T303^2]^4 & + \ldots = 0. & [30,12,24 ]
	\label{eq:T303_ideal}
\end{align}
Obviously, the relations for $\T220$, $\T121$, $\T221$, $\T321$, and $\T330$ can be easily obtained by utilizing the symmetry $\lamV \leftrightarrow \mV$. It is worth pointing that these relations can, in principle, be checked by substituting the invariants by their representation in terms of the parameters of the Higgs potential \eqref{eq:V_gen}. We use this approach to analytically cross check the relations of lower degrees (for $\T111$, and $\T202$) via \FORM or/and \Mathematica. For higher degrees this becomes very time-consuming and impractical due to a large number of terms that should eventually cancel. 
In practice, this issue can be addressed by using a numerical approach, i.e., by assigning random values to relevant parameters and checking the cancellation with the desired numerical precision. 
However, in this work we again use \Macaulay\ to prove that the relations belong to the ideal $\IdealTHDM$.\footnote{Since we have a syzygy expressing $\T303^2$ in terms of the even invariants \eqref{eq:syz_606} not involving $\mV$, it seems that the corresponding relation of the form \eqref{eq:algebric_relation} should be simple. However, one sees that the relation \eqref{eq:T303_ideal} has the highest total degree. This is due to the fact that we need to eliminate $\T202$ from \eqref{eq:syz_606} in favour of \eqref{eq:alg_indep_invs_unigraded}. The simplest way to do this is to consider the ideal generated by \eqref{eq:syz_606} and \eqref{eq:T202_ideal} alone and eliminate $\T202$.}

\bibliography{thdm_inv_6L.bib}

\end{document}